\documentclass[a4paper,10pt]{article}

\usepackage[utf8]{inputenc} 
\usepackage{mathtools,upgreek}
\usepackage{amssymb,amsmath,amsthm} 
\usepackage{graphicx} 
\usepackage[top=1.5cm,bottom=1.5cm,left=1.5cm,right=1.5cm]{geometry}
\usepackage{parskip} 
\usepackage[usenames,dvipsnames,svgnames,table]{xcolor}
\definecolor{cbOrange}{rgb}{0.9,0.6,0.0}
\definecolor{cbSkyBlue}{rgb}{0.35,0.7,0.9}
\definecolor{cbBluishGreen}{rgb}{0.0,0.6,0.5}
\definecolor{cbYellow}{rgb}{0.95,0.9,0.25}
\definecolor{cbBlue}{rgb}{0.0,0.45,0.7}
\definecolor{cbVermillion}{rgb}{0.8,0.4,0.0}
\definecolor{cbReddishPurple}{rgb}{0.8,0.6,0.7}

\renewcommand{\alpha}{\upalpha}
\renewcommand{\beta}{\upbeta}
\renewcommand{\gamma}{\upgamma}
\renewcommand{\delta}{\updelta}
\renewcommand{\psi}{\uppsi}
\renewcommand{\chi}{\upchi}
\renewcommand{\phi}{\upphi}
\renewcommand{\rho}{\uprho}

\renewcommand{\pi}{\uppi}
\renewcommand{\varphi}{\upvarphi}
\renewcommand{\theta}{\uptheta}
\renewcommand{\omega}{\upomega}
\renewcommand{\kappa}{\upkappa}

\renewcommand{\vec}[1]{\ensuremath{\mathbf{#1}}} 

\everymath{\mathtt{\xdef\tmp{\fam\the\fam\relax}\aftergroup\tmp}}

\newcommand{\ca}[1]{\textcolor{cbVermillion}{#1}}
\newcommand{\cb}[1]{\textcolor{cbOrange}{#1}}
\newcommand{\cc}[1]{\textcolor{cbBluishGreen}{#1}}
\newcommand{\cd}[1]{\textcolor{cbReddishPurple}{#1}}

\usepackage{units}
\usepackage{wasysym}
\usepackage{fancyhdr}

\DeclarePairedDelimiter\floor{\lfloor}{\rfloor}
\author{Axel Schild, Laboratory for Physical Chemistry, ETH Z\"urich, Switzerland}
\title{On the Probability Density of the Nuclei in a Vibrationally Excited Molecule}

\begin{document}

\maketitle

\begin{abstract}
  For localized and oriented vibrationally excited molecules, the 
  one-body probability density of the nuclei (one-nucleus density) is studied.
  Like the familiar and widely used one-electron density that represents the 
  probability of finding an electron at a given location in space, the 
  one-nucleus density represents the probability of finding a nucleus at a 
  given position in space independent of the location of the other nuclei.
  In contrast to the full many-dimensional nuclear probability density, the
  one-nucleus density contains less information and may thus be better 
  accessible by experiment, especially for large molecules.
  It also provides a quantum-mechanical view of molecular vibrations that can 
  easily be visualized.
  We study how the nodal structure of the wavefunctions of vibrationally 
  excited states translates to the one-nucleus density. 
  It is found that nodes are not necessarily visible:
  Already for relatively small molecules, only certain vibrational excitations 
  change the one-nucleus density qualitatively compared to the ground state.
  It turns out that there are some simple rules for predicting the shape of the 
  one-nucleus density from the normal mode coordinates, and thus for predicting 
  if a vibrational excitation is visible in a corresponding experiment.
\end{abstract}

Quantum-mechanically, the state of an approximately isolated molecule is
described by a wavefunction that depends on the location of all nuclei and 
electrons.
The corresponding probability density, which represents the distribution of the 
particles in the high-dimensional configuration space of their coordinates, is 
thus difficult to visualize, to comprehend, and also to measure.
Notwithstanding, there is an intuitive semi-classical picture of a molecule
in chemistry.
The emergence of this picture is a non-trivial problem\cite{sutcliffe05pccp,sutcliffe10tca} 
and looking at static states\cite{matyus11pra,matyus11jcp,matyus12jcp} as 
well as the quantum dynamics\cite{manz2014,bredtmann2015,torres2015,diestler2018}
of small isolated molecules can lead to surprising insights.
An important role for understanding the chemical picture of a molecule is 
certainly played by the large difference of electronic and nuclear masses.
This mass difference results in a strong spatial localization of the nuclei 
compared to the electrons, which in turn motivates a separation of the molecular 
wavefunction into a (marginal) nuclear wavefunction and an electronic 
wavefunction that conditionally depends on the location of the nuclei.
While such a separation is exact\cite{abedi12}, its practical application is 
usually in terms of the Born-Oppenheimer approximation \cite{born1927adp}
where the effect of the nuclear motion on the electronic wavefunction is 
neglected and only the chosen position of the nuclei is relevant \cite{schild2016}.

In the semi-classical picture of a molecule, the theoretical treatment of 
nuclei and electrons is different:
The delocalized electrons are considered to be quantum particles, while the 
comparably localized nuclei are often approximated as classical particles.
However, nuclei are also quantum particles and for small molecules the nuclear 
(many-body) probability densities have often been calculated and analyzed.\cite{smit2001,dawes13,welsch15,donoghue2016}
Also, there has been recent interest
in measuring e.g.\ the nuclear probability density \cite{shapiro1981,zewail00,jurek2004,ergler2006,schmidt2012,kimberg2014,zeller2016}
or the nuclear flux (current) density.\cite{manz13prl,barth15,bredtmann2015pccp}
An intriguing example of the quantum nature of the nuclei is the 
measurement of the nuclear density of a vibrationally excited H$_2^+$-molecules 
by means of Coulomb explosion imaging.\cite{schmidt2012}
The measured nuclear density, which depends only on the relative distance of 
the two nuclei, shows the expected nodal pattern of vibrationally excited states.

Similar measurements can be made for the electronic probability density, as 
exemplified by the measurement of the correlated two-electron probability 
density of the H$_2$ molecule.\cite{waitz2017}
For more than two electrons, however, there is a dimensionality problem
because multiple electrons have to be measured in coincidence and their 
probability density is difficult to grasp for the human intuition which is based 
on a three-dimensional experience of the world.
What can be done instead is to consider the electronic one-body probability 
density, also known as the one-electron density.
The one-electron density is the marginal density of finding one electron at a 
given location in space independent of where the other electrons are.
It is the central quantity of Density Functional Theory  \cite{ullrich2012}, it 
is easily visualized, and it is directly accessible to experiment if the 
molecule is localized:
For example, an electron scanning tunneling microscope does essentially measure 
the one-electron density of a localized molecule and can provide intuitive
images of molecules on surfaces \cite{moore2008}.
However, while some information about the electronic state can be extracted 
from the one-electron density,\cite{baer2010} this task is in general difficult:
Although the many-electron density is qualitatively different for different 
excited states due to the appearance of nodes in the wavefunction, the 
corresponding one-electron densities may be very similar.

Inspired by the experimental imaging of the one-electron density, the question 
arises if one-nucleus densities can be measured and what information about the 
nuclear state they contain.
To measure a one-nucleus density, only the position of one nucleus relative to 
the lab frame needs to be measured, hence visualization and interpretation of 
experimental data can be more direct than in coincidence measurements of the 
relative 
position of multiple nuclei.
Similar to the one-electron density, the one-nucleus density can also represent
a three-dimensional picture of all nuclei if the experiment is insensitive to 
the type of the nuclei or if the data of different types of nuclei are combined.
It can provide a straightforward visualization of the quantum state of the 
nuclei in a molecule and can in this way complement our understanding of 
molecular behavior.
However, to obtain the one-nucleus density the molecule needs to be localized, 
e.g.\ on a surface or with the help of a trap\cite{mccarron2018}.
This author is not aware of any experiments that provide the one-nucleus density
of a molecular system with an accuracy that can e.g.\ resolve 
vibrational excitations, but in principle such experiments are possible.

Clearly, a measurement of the one-nucleus density would provide a picture 
of the quantum nature of nuclei that is directly accessible to our spatial 
conception.
But would it provide information about the vibrational state of the molecule, 
i.e., would the nodal structure of wavefunctions in excited states be 
visible in the one-nucleus density?
This question is investigated in the following with the help of the nuclear
wavefunction obtained from a normal mode analysis, i.e., obtained from the 
local harmonic approximation of the potential energy surface for the nuclear 
configuration of lowest energy\cite{henriksen2011}.
The nuclear wavefunction $\psi_{\rm nuc}$ is then a product of a translational, 
a rotational, and a vibrational part,
\begin{align}
 \psi_{\rm nuc} = \psi_{\rm nuc}^{\rm trans} \times \psi_{\rm nuc}^{\rm rot} \times \psi_{\rm nuc}^{\rm vib}.
 \label{eq:stupid}
\end{align}
The translational part $\psi_{\rm nuc}^{\rm trans}$ represents translation of the whole 
molecule in space and can always be factored exactly, while the separation of
$\psi_{\rm nuc}^{\rm rot}$ (which describes rotations of the whole molecule) from 
$\psi_{\rm nuc}^{\rm vib}$ is only valid for small displacements 
from the equilibrium configuration\cite{eckart1935pra,littlejohn1997rmp,bunker2006,lauvergnat2016jcp}.
Typically, a normal mode analysis aims at computing the vibrational frequencies
(and maybe analyzing the normal mode coordinates) while $\psi_{\rm nuc}$ is of 
little interest.
However, if those frequencies are in good agreement with measured 
frequencies, the function $\psi_{\rm nuc}$ is likely also a good approximation 
to the exact nuclear wavefunction.
Then, statements about how qualitative features of $\psi_{\rm nuc}$ translate 
to the approximate one-nucleus density can be expected to be also true for the 
exact one-nucleus density.

In the following, the one-nucleus density of the nuclei for different states 
of $\psi_{\rm nuc}^{\rm vib}$ is studied and it is investigated how the nodes 
of the wavefunction in excited states manifest in the one-nucleus density.
As shown below, the nuclei are rather localized.
In analogy to the classical representation of the $N$ nuclei as $N$
points in a three-dimensional space, the sum of the one-nucleus densities for 
all individual (types of) nuclei yields a density in three-dimensional space 
where the probability distribution of each nucleus is clearly visible.
For brevity, hereafter this sum is simply called the one-nucleus density (in 
analogy to the one-electron density) and the nuclear many-body probability 
density in the $N$-dimensional configuration space is called the $N$-nucleus density.
A detailed description of how the one-nucleus densities are obtained is given 
in the Supporting Information.
Here, only the approximations and assumptions are discussed to 
point out when the approach is applicable.
The aim is to determine the one-nucleus density $\rho(\vec{R})$ from 
the approximate nuclear wavefunction $\psi_{\rm nuc}(X)$,
where $\vec{R} = (R_1,R_2,R_3)$ is a three-component vector and where 
$X = (\vec{X}_1, \dots, \vec{X}_{N})$
stands for the $N$ three-component position vectors $\vec{X}_j$ of the nuclei.
The one-nucleus densities for each nucleus are obtained by integrating the 
$N$-nucleus density $|\psi_{\rm nuc}(X)|^2$ over all but the coordinates of the selected nucleus,
\begin{align}
\rho_j(\vec{R}) = \left. \int \dotsi \int |\psi_{\rm nuc}(X)|^2 d\vec{X}_{\{1 \cdots N_n\} \setminus j} \right|_{\vec{X}_j=\vec{R}}
\label{eq:rhoJ}
\end{align}
for $d\vec{X}_{\{1 \cdots N_n\} \setminus j} = d\vec{X}_1 \dots d\vec{X}_{j-1} d\vec{X}_{j+1} \dots d\vec{X}_{N_n}$.
The one-nucleus density of the molecule is the sum of the one-nucleus densities
for all nuclei,
\begin{align}
\rho(\vec{R}) &= \sum_{j=1}^{N} \rho_j(\vec{R}).
\label{eq:rhosum}
\end{align}
The one-nucleus density gives the probability to find any nucleus in a given 
region of space, but from the relative spatial location it is immediately clear 
if the nucleus is e.g.\ an oxygen or a hydrogen nucleus.

The nuclear wavefunction is obtained as follows:
1) The Born-Oppenheimer approximation is made to obtain a Schr\"odinger equation 
for the nuclear wavefunction alone, with a potential energy surface $V(X)$.
Only the electronic ground state is considered.
2) A standard normal mode analysis\cite{henriksen2011} at one of 
the minima of $V$ is made.
From this calculation $N$ normal mode coordinates $q_j$ and the frequencies of 
their harmonic oscillator (HO) potentials are obtained.
The nuclear wavefunction is a product of HO wavefunctions in each normal mode 
coordinate, and coupling of rotational and vibrational degrees of freedom is 
neglected.\cite{littlejohn1997rmp,bunker2006}
3) There are six (five for linear molecules) normal modes for which the 
corresponding HO frequency is zero, representing translation and rotation of 
the molecule.
The nuclear wavefunction has the form of \eqref{eq:stupid}.
It is assumed that 
$\psi_{\rm nuc}^{\rm trans}$ and $\psi_{\rm nuc}^{\rm rot}$ are normalized 
Gaussian functions with a very small width corresponding to a frequency of 
\unit[0.5]{$E_h/\hbar$}.
This acts like a constraint on the coordinates and has the effect of 
localizing and orienting the molecule.
The resulting probability density can be interpreted either as a cut through the full 
density or as the nuclear density given the molecule is at a certain position 
and oriented in a certain way.
A practical advantage of this choice for $\psi_{\rm nuc}^{\rm trans}$ and 
$\psi_{\rm nuc}^{\rm rot}$ is that the wavefunction becomes a product of HO 
eigenfunctions in all modes.
The nuclear wavefunction is
\begin{align}
 \psi_{\rm nuc}(X) = \prod_{j=1}^{3 N} \phi_j\left(q_j(X),m_j\right),
 \label{eq:ho}
\end{align}
where  $\phi_j(q_j,m_j)$ is a HO wavefunction with quantum number $m_j$, and 
where $m_j = 0$ for the normal mode coordinates for translation and rotation of 
the whole molecule.
In the Supporting Information, it is described how the high-dimensional integral 
of \eqref{eq:rhoJ} with wavefunction \eqref{eq:ho} can be done analytically.
All quantum chemical calculations are made with the program Psi4\cite{turney12} 
using 3rd order M\o ller-Plesset pertubation theory and a cc-pVTZ basis 
set\cite{dunning1989jcp}.

\begin{figure*}[htbp]
  \centering
  \includegraphics[width=.99\textwidth]{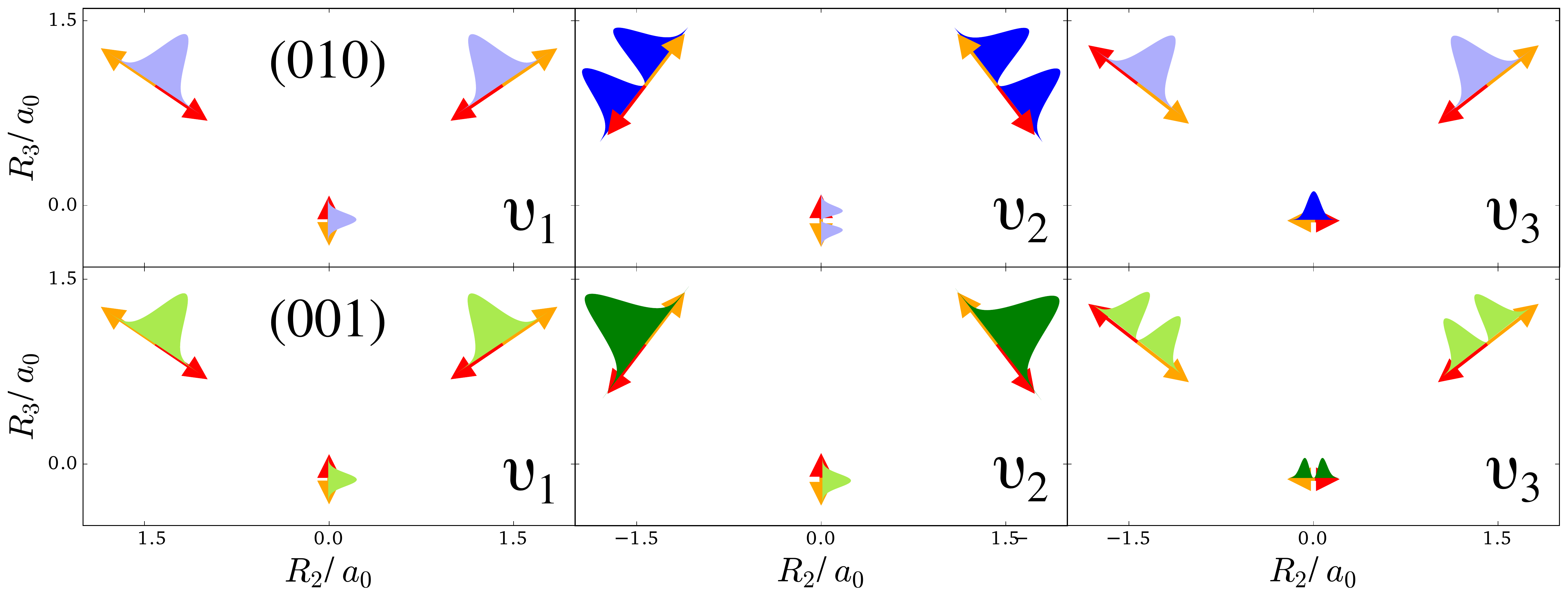}
  \caption{
  Normal modes of a water molecule (arrows showing the 
  extent and directionality (color) of the nuclear displacement along the mode):
  symmetric stretch $\nu_1$, bending mode $\nu_2$, and antisymmetric stretch $\nu_3$.
  The top-row shows sketches of the harmonic oscillator densities along the 
  modes for the first excitation of the symmetric stretch (010), the 
  bottom row for the first excitation of the antisymmetric stretch (001).
  A light filling of these densities means that locally other modes point in 
  the same direction, while a dark filling means that this is not the case.
  }
  \label{fig:pic_water_loco}
\end{figure*}

\begin{figure*}[htb]
  \centering
  \includegraphics[width=.99\textwidth]{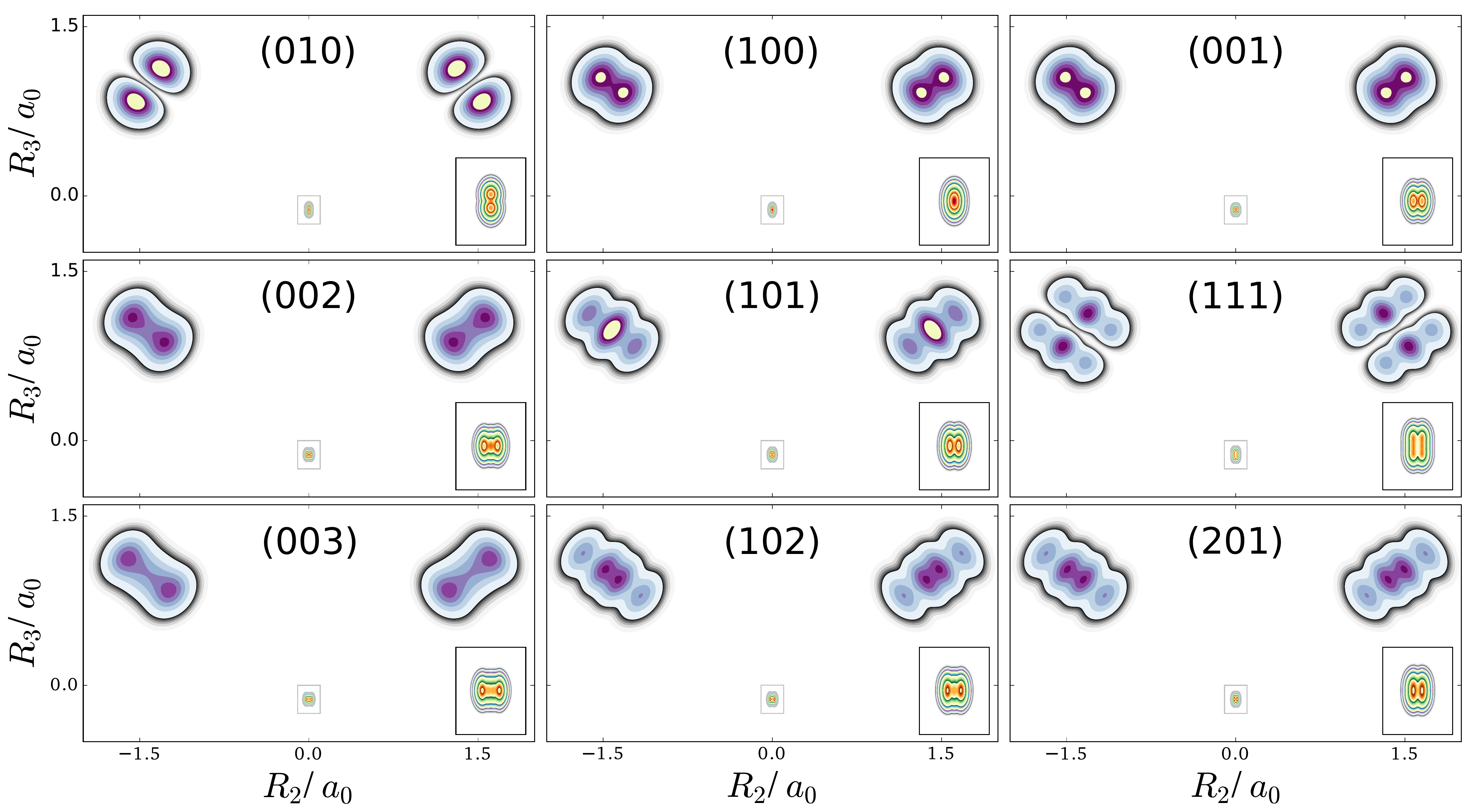}
  \caption{Contour plots of the one-nucleus density of a localized and 
  oriented water molecule in the molecular plane for different 
  vibrationally excited states.
  State labels $(m_1,m_2,m_3)$ indicate the number of quanta in normal modes 
  $\nu_1$, $\nu_2$, $\nu_3$, respectively.}
  \label{fig:pic_water_density}
\end{figure*}

The first example is the one-nucleus density of the water molecule.
Figure \ref{fig:pic_water_loco} shows its familiar vibrations as arrows 
indicating the motion of classical nuclei.
There are three vibrational normal mode coordinates which are labeled according 
to the usual spectroscopic notation\cite{shimanouchi1972}:
the symmetric stretch $\nu_1$, the bending mode (scissoring) $\nu_2$, and the 
antisymmetric stretch $\nu_3$.
The one-nucleus densities of the water molecule for some excitations of these 
modes are shown in Figure \ref{fig:pic_water_density} as contour plots in the 
molecular plane, labeled as $(m_1,m_2,m_3)$, where $m_j$ is the quantum number 
of mode $\nu_j$.
Insets magnify the details of the nuclear density around the oxygen nucleus,
as the density there is much more localized compared to the hydrogen nuclei
due to the mass difference.

The one-nucleus density of the vibrational ground state for each nucleus looks 
like a product of Gaussian functions oriented along the directions of the normal 
modes (not shown).
The one-nucleus densities of the first excitation in each mode, (100), (010), and 
(001), show how a node in the first excited state of the HO along the 
corresponding normal modes translates to the one-nucleus density.
From the pictures, it seems that the node in the wavefunction of one of 
these modes leads to a depletion of the one-nucleus density along this normal mode 
coordinate, but not to exact nodes or nodal planes in the one-nucleus density.
There are two reasons for the absence of exact nodes:
First, a Gaussian distribution for the translational and rotational modes is 
assumed. 
Depending on the width of the these distributions, the resulting one-nucleus 
densities become broader and loose their structure.
A very narrow Gaussian distribution is chosen, hence this reason is of minor 
importance.
The main reasons for the absence of exact nodes in the one-nucleus density is that 
those nodes only exists in configuration space, while the reduction of the 
$N$-nucleus density to the one-nucleus density as given in \eqref{eq:rhoJ} 
in general does not yield zero anywhere in space.

To understand the one-nucleus density in Figure \ref{fig:pic_water_density}, the 
analytic form of the $N$-nucleus density needs to be investigated.
It is a product of HO densities in all normal modes, because the wavefunction 
\eqref{eq:ho} is a product of HO wavefunctions. 
These 1d-HO densities can be visualized by functions centered at the 
equilibrium position of the three nuclei, with extent and direction as 
given by the arrows.
In Figure \ref{fig:pic_water_loco}, the idea is illustrated for the first
excited state of the bending mode, $(010)$, and of the antisymmetric stretch, 
$(001)$.

Some predictions can be made about the qualitative features of the one-nucleus densities
at the nuclei by means of a set of simple rules.
These rules are called the LOcal COmparison (LOCO) rules, because they are 
based on a comparison of the normal mode coordinates at the location of each 
nucleus separately.
The LOCO rules are as follows:
For a nucleus, the magnitudes and directions of the displacements along the 
normal modes are compared.
(a) If only one normal mode displaces the nucleus in a certain direction or if 
there is one normal mode that displaces the nucleus in a certain direction much 
stronger than the other normal modes, the nodes of the wavefunction due to an 
excitation of this normal mode are clearly visible as depletions in the 
one-nucleus density.
(b) If several normal modes displace the nucleus in the same direction by 
similar magnitude, an excitation of one of these modes is not necessarily 
visible in the one-nucleus density.
In general, the more such modes exist, the less likely it is that an excitation 
in one of these can be recognized in the one-nucleus density.
It follows that typically, only the normal modes that displace a nucleus the 
most in a given direction can have a strong influence on the qualitative shape 
of the one-nucleus density.
(c) If there are two (or more) modes that displace the nucleus in the same
direction, simultaneous excitation of these modes may show combination 
features, as exemplified below.

For example, the one-nucleus density of states (100), (010) and (001) can be 
understood from the LOCO rules (a) and (b) as follows:
In Figure \ref{fig:pic_water_loco}, sketches of the harmonic oscillator 
wavefunctions in the modes are shown.
At the hydrogen nuclei, $\nu_1$ and $\nu_3$ point in a similar direction, while
$\nu_2$ is perpendicular.
Thus, the excitation of $\nu_2$ (state (010)) leads almost to a nodal plane 
in the one-nucleus density at the hydrogen nuclei, cf.\ Figure \ref{fig:pic_water_density}.
In contrast, excitation of $\nu_1$ (state (100)) or $\nu_3$ (state (001)) lead to a 
significantly less pronounced depletion of the one-nucleus density at the equilibrium 
position of hydrogen.
For the oxygen nucleus, the situation is reversed, as $\nu_1$ and $\nu_2$ point 
in the same direction and $\nu_3$ is perpendicular.
Consequently, the one-nucleus density of state (001) shows a depletion at the oxygen 
equilibrium position, while no depletion is seen in state (100).
As $\nu_2$ displaces the oxygen nucleus stronger than $\nu_1$ (which is, however, 
hardly visible on the scale of Figure \ref{fig:pic_water_loco}), a depletion 
due to the node in $\nu_2$ is visible in state (010).

A situation similar to that of state (001) is found for states (002) and (003), 
i.e.\ when the HO function of $\nu_3$ is in its second or third excited state.
For (002), there is the expected triple-maximum structure at the oxygen nucleus 
(with a lower central maximum), while the central maximum at the hydrogen nuclei 
is not visible.
For state (003) four maxima are found in the one-nucleus density at the oxygen 
nucleus (although the two central ones are too weak to be clearly seen in Figure 
\ref{fig:pic_water_density}), while at the hydrogen nuclei still only two 
maxima can be found.

\begin{figure*}[htb]
  \centering
  \includegraphics[width=.99\textwidth]{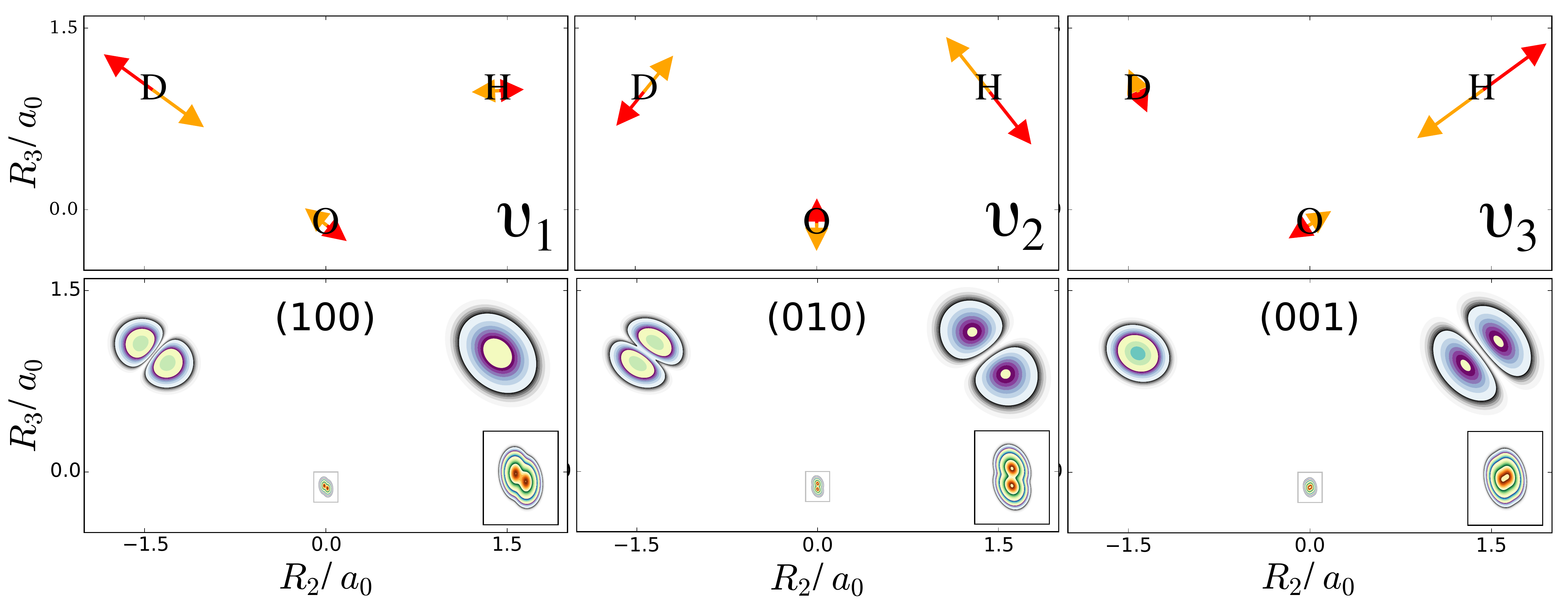}
  \caption{Top: Normal mode coordinates of the mono-deuterated water molecule:
  O-D stretch $\nu_1$, bending mode $\nu_2$, O-H stretch $\nu_3$.
  Bottom: Contour plots of the one-nucleus density of a localized 
  and oriented mono-deuterated water molecule in the molecular plane for 
  vibrational states corresponding to the first excitations of the normal modes 
  shown above.
  }
  \label{fig:pic_dwater}
\end{figure*}

An example for LOCO rule (c) is found if two normal modes are excited that 
locally have similar magnitude and direction.
For state (101) three maxima appear at the hydrogen nuclei, similar to a 
second excited state of the HO, but the central maximum is strongest while the 
outer maxima are weaker.
At the oxygen nucleus only two maxima that point along the direction of 
normal $\nu_3$ are found, i.e.\ a combination of the features of the one-nucleus 
densities for states (100) (only one maximum in the direction of $\nu_1$) and 
(001) (two maxima in the direction of $\nu_3$).
For states (102) and (201) the qualitative features of the one-nucleus densities 
can be explained analogously and in accord with the LOCO rules, especially the 
four maxima at the hydrogen nuclei and a triple maximum at the oxygen nucleus 
for state (102), but only a double maximum at the state (201).

Last, for state (111) the resulting one-nucleus density is a simple combination
of features of states (010) and (101), because $\nu_2$ is locally perpendicular 
to the other two vibrational modes at the hydrogen nuclei.
At the oxygen nucleus, the effect of exciting two modes with similar
direction and magnitude, LOCO rule (c), is that three maxima are found along 
coordinate $R_3$ upon closer inspection, although the central one is hardly 
visible.

The normal modes of the water molecule have a symmetry with respect to the 
hydrogen nuclei.
If the symmetry of the nuclear structure is broken by replacing one hydrogen 
nucleus with a deuterium nucleus, a very different picture for the 
one-nucleus densities is obtained.
The normal mode coordinates for a mono-deuterated water molecule are given in 
Figure \ref{fig:pic_dwater}.
The bending mode $\nu_2$ is similar to the bending mode of water, but $\nu_1$ 
is now the O-D stretch mode that only displaces the deuterium nucleus strongly, 
while $\nu_3$ is the O-H stretch mode that almost exclusively displaces the 
hydrogen nucleus.
Thus, according to the LOCO rules it is expected that any excitation of $\nu_2$ 
is clearly visible in the one-nucleus density at the hydrogen and deuterium nucleus, 
that an excitation of $\nu_1$ is only visible at the deuterium nucleus, and that 
and excitation of $\nu_3$ is only visible at the hydrogen nucleus.
The prediction of the LOCO rules is accurate, as the one-nucleus densities of 
the first excited states of each of these modes shown in Figure 
\ref{fig:pic_dwater} illustrate.

A consequence of the LOCO rules is that for molecules with many nuclei, only 
certain (of the lowest) excited states are qualitative visible in the 
one-nucleus density, i.e.\ those that are the ones displacing a nucleus the most 
in a certain direction of space.
This is indeed the case.
For example, for the benzene molecule the one-nucleus density of none of the 
first excited states of any normal mode in the molecular plane is qualitatively 
different from the vibrational ground state.
The situation changes if one hydrogen nucleus is exchanged by a deuterium 
nucleus.
In the Supporting Information, the one-nucleus densities for the only two normal 
mode coordinates that strongly displace the deuterium nucleus are shown. 
Further examples given in the Supporting Information are an analysis of 
ethene and mono-deuterated ethene as molecules that contain less nuclei than 
benzene but where similar effects are found, and methane as an example of a 
non-planar molecule.
The TOC graphic shows the one-nucleus density of one of the vibrational states 
responsible for the blue color of water.\cite{braun1993}

In conclusion, it is found that similar to how a one-electron density provides
understanding of the behavior of electrons, the one-nucleus densities of 
molecules can provide an interesting opportunity to measure and to visualize 
the quantum nature of the nuclei:
Compared to the $N$-nucleus density, where the location of all $N$ nuclei 
of a molecule needs to be known, the one-nucleus density provides an accessible 
representation of the nuclear structure that complements the classical picture 
of the nuclei.
While a measurement of the one-nucleus density is undoubtedly difficult because 
the nuclei are localized in a relatively small region, it is in principle 
possible, especially for light ``quantum'' nuclei like hydrogen or deuterium
that are comparably delocalized.

Importantly, in contrast to the one-electron density where the electronic state 
is in general not qualitatively visible (or where, to this authors knowledge,
there are no rules available to predict if this is the case), the vibrational 
state of a molecule may be directly visible in the one-nucleus density.
The presented LOCO rules allow to predict, from a normal mode analysis of the 
nuclear wavefunction, without much effort which vibrational excitations would 
be visible in the one-nucleus density and thus which molecules are rewarding 
targets for experimental investigations.

{\bf Acknowledgement}

The author is grateful to J.\ Manz (Freie Universit\"at Berlin) for 
stimulating and supporting this work and to E.K.U.\ Gross 
(MPI $\mu\Phi$ Halle) for extensive discussions about the topic.

{\bf Supporting Information Available:} Details of the computation and further 
examples of one-nucleus densities.

\clearpage

\begin{center}
  \Large Supporting Information for ``On the Probability Density of the Nuclei in a Vibrationally Excited Molecule'':
Computational Details
\end{center}

\section{The local harmonic approximation}

The local harmonic approximation is briefly reviewed.
For a detailed description, see \cite{henriksen2011}.
In the Born-Oppenheimer approximation, the nuclear wavefunction is obtained 
from
\begin{align}
 \left(-\sum_{j=1}^N \frac{\hbar^2 \partial_{X_j}^2}{2 M_j} + V(X) \right) \Psi(X) &= E \Psi(X)
 \label{eq:se}
\end{align}
with masses $M_j$ for the coordinates $X_j$.
Around a minimum $X_{eq}$ of $V$, the potential is expanded in a Taylor series 
to second order in terms of the displacement coordinates $x = X-X_{eq}$ and 
$V(X_{eq})$ is set to zero.
Additionally, mass-weighted displacement coordinates $\tilde{x}_j = 
\sqrt{M_j} x_j$ are introduced.
Then \eqref{eq:se} becomes
\begin{align}
 \left(-\sum_{j=1}^N \frac{\hbar^2 \partial_{\tilde{x}_j}^2}{2} + 
       \frac{1}{2} \sum_{j=1}^N \sum_{k=1}^N U_{jk} \tilde{x}_j \tilde{x}_k
       \right) \Psi(X(\tilde{x})) &= E \Psi(X(\tilde{x}))
\end{align}
with
\begin{align}
 U_{jk} =\frac{\left.\partial_{X_j} \partial_{X_k} V(X) \right|_{X=X_{eq}}}{\sqrt{M_j M_k}}
\end{align}
The matrix $U$ is diagonalized by the matrix of eigenvectors $Q$, 
\begin{align}
 Q^T \cdot U \cdot Q = \Omega
\end{align}
where $Q^T Q = Q Q^T = diag(1)$ and  $\Omega = diag(\omega^2)$ is the diagonal 
matrix of the eigenvalues $\omega_j^2$.
This yields
\begin{align}
 \sum_{j=1}^N \sum_{k=1}^N U_{jk} \tilde{x}_j \tilde{x}_k 
  &= \tilde{x}^T \cdot U \cdot \tilde{x} 
   = \tilde{x}^T \cdot Q \cdot Q^T U  \cdot Q \cdot Q^T \cdot \tilde{x} 
   = \sum_{j=1}^N \omega_j^2 \tilde{q}_j^2
\end{align}
with normal mode coordinates
\begin{align}
 \tilde{q}_j = Q^T \tilde{x} = \sum_{k=1}^N Q_{kj} \tilde{x}_k = \sum_{k=1}^N \sqrt{M_k} Q_{kj} x_k.
\end{align}
In terms of the normal mode coordinates, \eqref{eq:se} becomes
\begin{align}
 \left(\sum_{j=1}^N \left(-\frac{\hbar^2 \partial_{\tilde{q}_j}^2}{2} + 
        \frac{\omega_j^2}{2} \tilde{q}_j^2 \right)
       \right) \Psi(X(\tilde{q})) &= E \Psi(X(\tilde{q}))
\end{align}
With the coordinate scaling
\begin{align}
 q_j &= \sqrt{\frac{\omega_j}{\hbar}} \tilde{q}_j 
  = \sum_{k=1}^N \sqrt{\frac{\omega_j M_k}{\hbar}} Q_{kj} x_k
 \label{eq:xtoq}
\end{align}
the wavefunction becomes a product
\begin{align}
 \Psi(X(q)) := \psi(q) = \prod_{j=1}^N \phi_j(q_j,m_j)
 \label{eq:product}
\end{align}
with the eigenfunctions of the quantum harmonic oscillator with unit mass
and unit frequency\cite{cohentannoudji2007,jeffrey2008}
\begin{align}
 \phi_j(q_j,m_j) &= \sqrt{2^{m_j} m_j!}
                    \left(\frac{1}{\pi} \right)^{\frac{1}{4}}
                    e^{-\frac{q_j^2}{2}} 
                    \sum_{k=0}^{\floor*{\frac{m_j}{2}}} \frac{(-1)^k}{4^k k! (m_j-2k)!} 
                    q_j^{m_j-2k},
\end{align}
where the floor function $\floor{r}$ gives the largest integer smaller than or 
equal to $r$.
However, there is one catch:
Six (or five, for linear molecules) of the frequencies $\omega_j$ are zero, as
they belong to translations and rotations of the full system.
There are no external potentials that break the translational and rotational
invariance of the nuclear system, hence it is broken artificially by setting a
non-zero value for these frequencies $\omega_j$.
The quantum numbers $m_j$ for these degrees of freedom are set to zero so that 
the density in these modes is a Gaussian function with a width determined by 
the chosen frequency.
In the limit $\omega_j \rightarrow 0$, a $\delta$-distribution is obtained
for $|\phi_j(q_j,0)|^2$.

\section{The density}

In the main article, the nuclear one-nucleus density is denoted as $\rho(\vec{R})$ and 
it is defined as sum of the individual one-nucleus densities $\rho_j(\vec{R})$.
In this notes it is described how to obtain $\rho_j(\vec{R})$, but the notation 
is slightly different.
We start from the density $\rho^{[N]}(x_1,\dots,x_N)$ in the $N$-dimensional 
configuration space in terms of displacement coordinates $x_j$, and we aim at 
computing the one-nucleus density $\rho^{[3]}(x_1,x_2,x_3)$ of the particle
with displacement coordinates $x_1,x_2,x_3$, by integrating over $x_4,\dots,x_N$.
To obtain $\rho(\vec{R})$, all we need to do is to shift $\rho^{[3]}(x_1,x_2,x_3)$
to the equilibrium position of the considered nucleus, to repeat the 
procedure for the coordinates of all other nuclei, and to add all those 
densities.
We use the notation $\rho^{[N]}$ because we give the solution of the integrals 
iteratively, by computing $\rho^{[N-1]}$, $\rho^{[N-2]}$, etc., with each of 
the densities in this series depending on one coordinate less compared to the 
previous density.

The explicit expression for the $N_n$-body or $N$-coordinate density 
$\rho^{[N]}(x)$ can now be given.
We require that 
\begin{align}
  \int \dotsi \int |\Psi(X)|^2 dX_1 \dots dX_N 
  = \int \dotsi \int \rho^{[N]}(x) dX_1 \dots dX_N 
  \stackrel{!}{=} 1
\end{align}
and have
\begin{align}
 \int |\phi_j(q_j,m_j)|^2 dq_j = 1,
\end{align}
where $\int$ represents the definite integral $\int_{-\infty}^{\infty}$ 
throughout this text.
Thus, the density is
\begin{align}
 \rho^{[N]}(x) &= J_{qx} \prod_{j=1}^N \left(\phi_j(q_j,m_j)\right)^2
\end{align}
with the Jacobian determinant $J_{qx}$ for the coordinate
transformation \eqref{eq:xtoq} from $x$ to $q$ that ensures that $\rho^{[N]}(x)$ 
is normalized to one when integrating over all $x$.
Explicitly, with the transformation matrix
\begin{align}
 T_{jk} = \sqrt{\frac{\omega_j M_k}{\hbar}} Q_{kj}
 \label{eq:transform}
\end{align}
we have
\begin{align}
 J_{qx} = |det(T_{jk})|
 \label{eq:jacob}
\end{align}
Then
\begin{align}
 \rho^{[N]}(x) &= J_{qx} \prod_{j=1}^N 
                    \frac{2^{m_j} m_j!}{\sqrt{\pi}} 
                    e^{-q_j^2} 
                    \sum_{k=0}^{\floor*{\frac{m_j}{2}}} \sum_{l=0}^{\floor*{\frac{m_j}{2}}}
                    \frac{(-1)^{k+l}}{4^{k+l} k! l! (m_j-2k)! (m_j-2l)!} q^{2(m_j-(k+l))} \\
               &= J_{qx} \Gamma^{[N]}(x) \prod_{j=1}^N P_j^{(2m_j)}(x)
 \label{eq:rho}
\end{align}
with the Gaussian function\footnote{
  A factor $\pi^{N/2}$ could be added to have $\Gamma^{[N]}(x)$ properly 
  normalized. Instead, it is included in the definition of the polynomial 
  coefficients.}
\begin{align}
 \Gamma^{[N]}(x) &= exp\left(-\sum_{j=1}^N q_j(x)^2\right)
 \label{eq:gamma}
\end{align}
and with the polynomial of order $2m_j$
\begin{align}
 P_j^{(2m_j)}(x) &= \frac{2^{m_j} m_j!}{\sqrt{\pi}}
                    \sum_{k=0}^{\floor*{\frac{m_j}{2}}} \sum_{l=0}^{\floor*{\frac{m_j}{2}}}
                    \frac{(-1)^{k+l}}{4^{k+l} k! l! (m_j-2k)! (m_j-2l)!} q(x)^{2(m_j-(k+l))}
\end{align}
Before inserting the explicit definition of $q_j(x)$, we rewrite this 
polynomial by changing the summation variables to $k+l \rightarrow k$,
$(k-l)/2 \rightarrow l$, so that 
\begin{align}
 P_j^{(2m_j)}(x) &= \sum_{k=0}^{2 \floor*{\frac{m_j}{2}}} c_{m_j,k} q(x)^{2(m_j-k)}
 = \sum_{k=0}^{2 \floor*{\frac{m_j}{2}}} p_j^{(2(m_j-k))}
 \label{eq:rho_mult}
\end{align}
with coefficients
\begin{align}
 c_{m_j,k} = \frac{2^{m_j-2k} m_j!}{\sqrt{\pi}} \sum_{l=-L_k}^{L_k}
                    \frac{(-1)^{k}}{
                    \left(\frac{k}{2}+l\right)! 
                    \left(\frac{k}{2}-l\right)! 
                    \left(m_j-2\left(\frac{k}{2}+l\right)\right)!
                    \left(m_j-2\left(\frac{k}{2}-l\right)\right)!}
\end{align}
that are obtained with summation boundaries
\begin{align}
 L_k = \frac{1}{2} \left(\floor*{\frac{m_j}{2}} - \left|k - \floor*{\frac{m_j}{2}} \right| \right)
\end{align}

\fbox{\parbox{\textwidth}{\footnotesize
An alternative that is more practical for numerical implementations is to 
change the summation variables to $k+l \rightarrow k$, $(k-l)/2 \rightarrow l$,
so that the coefficients become
\begin{align}
 c_{m_j,k} &= \frac{2^{m_j-2k} m_j!}{\sqrt{\pi}} \sum_{\substack{l=-L_k \\ 2|l}}^{L_k}
                    \frac{(-1)^{k}}{
                    \left(\frac{k+l}{2}\right)! 
                    \left(\frac{k-l}{2}\right)! 
                    \left(m_j-\left(k+l\right)\right)!
                    \left(m_j-\left(k-l\right)\right)!} \\
 L_k &= \floor*{\frac{m_j}{2}} - \left|k - \floor*{\frac{m_j}{2}} \right| 
\end{align}
where the sum over $l$ now has increments $\Delta l = 2$.
}}

\section{Initial parameters}

The general form of the $N$-nucleus density \eqref{eq:rho} is
\begin{align}
 \rho^{[N]}(x) &=  \Gamma^{[N]}(x)
                  \left( 
                  A^{[N](0)} + 
                  \sum_{k_1,k_2=1}^N A_{k_1 k_2}^{[N](2)} x_{k_1} x_{k_2} + 
                  \dots +
                  \sum_{k_1,\dots,k_{2M}=1}^N A_{k_1 \dots k_{2M}}^{[N](2M)} x_{k_1} \dots x_{k_{2M}}
                  \right)
 \label{eq:rhon}
\end{align}
with the Gaussian function \eqref{eq:gamma} given as
\begin{align}
 \Gamma^{[N]}(x) &= exp\left(-\sum_{j=1}^N \sum_{k=1}^N S_{jk}^{[N]} x_j x_k \right)
 \label{eq:gamma2}
\end{align}

It is a multivariate Gaussian function multiplied by a polynomial composed of 
monomials of degree $0, 2, \dots, 2M$, where
\begin{align}
 M = \sum_{j=1}^N m_j
\end{align}
is the sum of quantum numbers.
The superscript $[N]$ of the coefficients for the monomials $A^{[N](2\alpha)}$ 
and $S_{jk}^{[N]}$ indicates that those are the parameters of the $N$-nucleus
density.
Below, we see that when we integrate over coordinate $x_N$ we obtain an 
$N-1$-nucleus density $\rho^{[N-1]}$ that looks like \eqref{eq:rhon}, except 
that the sums terminate at $N-1$ and the coefficients changed.
By determining the new coefficients, we can  perform all integrals that are 
necessary to derive the one-nucleus density iteratively.
Note that the Jacobian determinant is included in the initial coefficients,
cf. \eqref{eq:coefffin}.

However, first we have to determine the initial values of the coefficients.
Inserting the transformation equation of the coordinates 
\eqref{eq:xtoq},\eqref{eq:transform} into the definition of the Gaussian 
function \eqref{eq:gamma} yields
\begin{align}
 S_{jk}^{[N]} = \sum_{l=1}^N T_{lj} T_{lk}
 \label{eq:s_initial}
\end{align}
Similarly, inserting \eqref{eq:xtoq},\eqref{eq:transform} into the definition
of the polynomials \eqref{eq:rho_mult} for each quantum number $j$ shows that 
these polynomials are a sum (over $k_j$) of monomials
\begin{align}
 p_j^{2(m_j-k_j)} &= J_{qx} \sum_{l_1=1}^N \dotsi \sum_{l_{2(m_j-k_j)}=1}^N
                     A_{l_1 \dots l_{2(m_j-k_j)}}^{(2(m_j-k_j)),j}
                     x_{l_1} \dots x_{l_{2(m_j-k_j)}}
\end{align}
with coefficients\footnote{These coefficients are symmetric w.r.t.\ exchange
of any two indices.}
\begin{align}
 A_{l_1 \dots l_{2(m_j-k_j)}}^{(2(m_j-k_j)),j} &= 
   c_{m_j,k_j} T_{j l_1} \dots T_{j l_{2(m_j-k_j)}}
 \label{eq:acoeff}
\end{align}
The polynomial occurring in the definition of the density \eqref{eq:rho} is
\begin{align}
 \prod_{j=1}^N P_j^{(2m_j)} &= \sum_{k_1=0}^{2\floor*{\frac{m_1}{2}}} \dotsi
                               \sum_{k_N=0}^{2\floor*{\frac{m_N}{2}}} 
                               p_1^{2(m_1-k_1)} \dots p_N^{2(m_N-k_N)}
 \label{eq:poly}
\end{align}
By comparing the definition of the coefficients in the density \eqref{eq:rhon} 
with the from of the polynomial \eqref{eq:poly} we see how to obtain the 
coefficients $A^{[N](2\alpha)}$:
First, we compute all monomial coefficients \eqref{eq:acoeff}.
Then, we take the tensor product along $j$,\footnote{These coefficients are 
not symmetric w.r.t.\ exchange of any two indices anymore, but only within 
certain blocks. This makes the equations later a little bit more elaborate.}
\begin{align}
 J_{qx} \times
 A_{l_1 \dots l_{2(m_1-k_1)}}^{(2(m_1-k_1)),1} \otimes
 A_{l_1 \dots l_{2(m_2-k_2)}}^{(2(m_2-k_2)),2} \otimes
 \dots  \otimes
 A_{l_1 \dots l_{2(m_N-k_N)}}^{(2(m_N-k_N)),N}
 \label{eq:coefffin}
\end{align}
for all possible combinations of the $k_j$-index.
The number of indices of the resulting object is the sum of the number of 
indices of the individual coefficients and corresponds to the order of the 
polynomial $(2\alpha)$ to which it belongs.
Adding all the results of the same order yields the initial coefficients 
$A^{[N](2\alpha)}$ of the $N$-nucleus density.

\section{Integration}

Next, we need to integrate over one variable, say, $x_N$. 
For this purpose, we need the integral\footnote{$\int$ is still the definite 
integral from $x=-\infty$ to $x=\infty$.} \cite{jeffrey2008}
\begin{align}
 \int e^{-ax^2 +bx + c} dx = \sqrt{\frac{\pi}{a}} e^{\frac{b^2}{4a}+c} = I_0
 \label{eq:i0a}
\end{align}
Taking the derivative of \eqref{eq:i0a} w.r.t.\ $b$ and comparing with the 
definition of the Hermite polynomials yields
\begin{align}
 \int x^n e^{-ax^2 +bx + c} dx = 
  \sum_{m=0}^{\floor*{\frac{n}{2}}} \frac{n!}{2^n m! (n-2m)!} \frac{b^{n-2m}}{a^{n-m}} I_0
  \label{eq:ima}
\end{align}

Integrating the density \eqref{eq:rhon} over $x_M$ yields
\begin{align}
 \rho^{[N-1]}(x) &= \int \rho^{[N]}(x) 
                  = \sum_{i=0}^M \sum_{j=0}^{2i}
                    \sum_{k_1=1}^{N-1} \dotsi \sum_{k_{i-j}=1}^{N-1}
                    \hat{P}_j^{(2i)} A_{k_1 \dots k_{2i-j} N \dots}^{[N](2i)} x_{k_1} \dots x_{k_{2i-j}} I_j^{[N]}
 \label{eq:rhonm1}
\end{align}
Here, $A_{k_1 \dots k_{2i-j} N \dots}^{[N](2i)}$ represents the coefficient
for the monomial of order $2i$ with $2i-j$ indices that run from $1$ to $N-1$, 
and the remaining $j$ indices set to $N$. 
The operator $\hat{P}_j^{(2i)}$ constructs the sum of all $\binom{2i}{j}$ 
permutations of the indices set to $N$ with those that run from $1$ to $N-1$.

\fbox{\parbox{\textwidth}{\footnotesize 
 To perform the integral of the density over the last coordinate $x_N$, we 
 first have to group the monomials in \eqref{eq:rhon} for each order according 
 to the exponent of $x_N$ (which corresponds to the index $j$ of $I_j$) to find
 \begin{align}
   \int \rho^{[N]}(x) =& 
      A^{[N](0)} I_0 
    + \sum_{k_1=1}^{N-1} \sum_{k_2=1}^{N-1} A_{k_1 k_2}^{[N](2)} x_{k_1} x_{k_2} I_0
    + \sum_{k_1=1}^{N-1} \left( A_{k_1 N}^{[N](2)} + A_{N k_1}^{[N](2)}\right) x_{k_1} I_1
    + A_{NN}^{[N](2)} I_2 \nonumber \\
    &
    + \sum_{k_1=1}^{N-1} \sum_{k_2=1}^{N-1} \sum_{k_3=1}^{N-1} \sum_{k_4=1}^{N-1} 
       A_{k_1 k_2 k_3 k_4}^{[N](4)} x_{k_1} x_{k_2} x_{k_3} x_{k_4} I_0 \nonumber \\
    &
    + \sum_{k_1=1}^{N-1} \sum_{k_2=1}^{N-1} \sum_{k_3=1}^{N-1}
       \left(A_{k_1 k_2 k_3 N}^{[N](4)} 
           + A_{k_1 k_2 N k_3}^{[N](4)} 
           + A_{k_1 N k_2 k_3}^{[N](4)} 
           + A_{N k_1 k_2 k_3}^{[N](4)} \right) x_{k_1} x_{k_2} x_{k_3} I_1 \nonumber \\
    &
    + \sum_{k_1=1}^{N-1} \sum_{k_2=1}^{N-1}
       \left(A_{k_1 k_2 N N}^{[N](4)} 
           + A_{k_1 N k_2 N}^{[N](4)} 
           + A_{N k_1 k_2 N}^{[N](4)} 
           + A_{k_1 N N k_2}^{[N](4)} 
           + A_{N k_1 N k_2}^{[N](4)} 
           + A_{N N k_1 k_2}^{[N](4)} \right) x_{k_1} x_{k_2} I_2 \nonumber \\
    &
    + \sum_{k_1=1}^{N-1}
       \left(A_{k_1 N N N}^{[N](4)} 
           + A_{N k_1 N N}^{[N](4)} 
           + A_{N N k_1 N}^{[N](4)} 
           + A_{N N N k_1}^{[N](4)} \right) x_{k_1} I_3
    + A_{N N N N}^{[N](4)} I_4 + \dots
 \end{align}
 At each order $2i$, the coefficients of each integral $I_j$ are obtained as 
 sum of the $\binom{2i}{j}$ possible permutations of the index $N$ occurring 
 $j$ times in the coefficients $A^{[N](2i)}$. Unfortunately, 
 in general $A^{[N](2i)}$ is not symmetric when exchanging two indices.
 However, it is constructed as tensor product of arrays that have this 
 property, hence this symmetry may be exploited to some extent, if desired.
}}

We now assume (correctly) that $\rho^{[N-1]}$ has the same functional 
form as $\rho^{[N]}$, but with new coefficients $A^{[N-1]}$ and $S^{[N-1]}$, 
\begin{align}
 \rho^{[N-1]}(x)  &\stackrel{!}{=} 
                  exp\left(-\sum_{j=1}^{N-1} \sum_{k=1}^{N-1} S_{jk}^{[N-1]} x_j x_k \right)
                  \left( 
                  A^{[N-1](0)} + 
                  \sum_{k_1,k_2=1}^{N-1} A_{k_1 k_2}^{[N-1](2)} x_{k_1} x_{k_2} + 
                  \dots +
                  \sum_{k_1,\dots,k_{2M}=1}^{N-1} A_{k_1 \dots k_{2M}}^{[N-1](2M)} x_{k_1} \dots x_{k_{2M}}
                  \right) 
\end{align}
The integrals $I_j^{[N]}$ that occur here are of the form \eqref{eq:ima} and 
are given by
\begin{align}
 I_0^{[N]} &= \sqrt{\frac{\pi}{S_{NN}^{[N]}}} 
               exp\left( -\sum_{j=1}^{N-1} \sum_{k=1}^{N-1} 
                \left( S_{jk}^{[N]} - \frac{S_{jN}^{[N]} S_{kN}^{[N]}}{S_{NN}^{[N]}}
                 \right) x_j x_k \right)
 \label{eq:i0}
\end{align}
and
\begin{align}
 I_n^{[N]} &= \sum_{m=0}^{\floor*{\frac{n}{2}}}
               \frac{(-1)^n n!}{4^m m! (n-2m)!} 
                \frac{I_0^{[N]}}{\left(S_{NN}^{[N]}\right)^{n-m}} 
                 \sum_{k_1=1}^{N-1} \dotsi \sum_{k_{n-2m}=1}^{N-1} 
                  S_{k_1 N}^{[N]} \dots S_{k_{n-2m} N}^{[N]}
                   x_{k_1}\dots x_{k_{n-2m}}
 \label{eq:iN}
\end{align}
respectively.

\fbox{\parbox{\textwidth}{\footnotesize
 Equations \eqref{eq:i0} and \eqref{eq:iN} can be derived from \eqref{eq:i0a}
 and \eqref{eq:ima} by making the identifications
 \begin{align}
  a = S_{NN}^{[N]} && 
  b = -2 \sum_{j=1}^{N-1} S_{jN}^{[N]} x_j && 
  c = -\sum_{j=1}^{N-1} \sum_{k=1}^{N-1} S_{jk}^{[N]} x_j x_k
 \end{align}
}}

We see from \eqref{eq:i0} that because $I_0^{[N]}$ occurs in all integrals
$I_n^{[N]}$, the new coefficients for the exponential are 
\begin{align}
 S_{jk}^{[N-1]} = S_{jk}^{[N]} - \frac{S_{jN}^{[N]} S_{kN}^{[N]}}{S_{NN}^{[N]}}
 \label{eq:siter}
\end{align}
The new coefficients for the polynomial are
\begin{align}
 A_{k_1 \dots k_{2\alpha}}^{[N-1](2\alpha)} &=
  \sqrt{\frac{\pi}{S_{NN}^{[N]}}} \sum_{i=\alpha}^M \sum_{j=2(i-\alpha)}^{2i}
   \frac{(-1)^j j!}{4^{i-\alpha} (i-\alpha)! (j-2(i-\alpha))!} 
    \frac{\hat{P}_j^{(2i)} A_{k_1 \dots k_{2i-j} N \dots}^{[N](2i)} S_{k_{2i-j+1}N}^{[N]} \dots S_{k_{2\alpha} N}^{[N]}}{\left(S_{NN}^{[N]}\right)^{j+\alpha-i}}
 \label{eq:aiter}    
\end{align}
Some remarks are in order:
First, we note that the indices $k_1, \dots$ now only have $N-1$ entries.
Second, the term $A_{k_1 \dots k_{2i-j} N \dots}^{[N](2i)} S_{k_{2i-j+1}N}^{[N]} 
\dots S_{k_{2\alpha} N}^{[N]}$ of the last equation has to be read as follows:
We take $A_{k_1 \dots k_{2i}}^{[N](2i)}$ and set the last $j$ indices equal to 
$N$.
Then, we make a tensor multiplication with so many vectors $S_{k_j N}^{[N]}$ 
that the resulting object has $2 \alpha$ indices.

\fbox{\parbox{\textwidth}{\footnotesize
 The second step of the integration of $\rho^{[N]}(x)$ over $x_N$ is to group 
 the resulting polynomial according to the orders of the monomials and add the 
 respective contributions.
 The structure of $\rho^{[N-1]}(x)$ can be visualized as follows:
 \begin{align}
  & \binom{0}{0} [0] \oplus [\ca{0_0}] \tag{i=0} \\
  & \binom{2}{0} [2] \oplus [\cb{0_0}]
  + \binom{2}{1} [1] \oplus [\cb{1_0}]
  + \binom{2}{2} [0] \oplus [\ca{0_1},\cb{2_0}] \tag{i=1} \\
  & \binom{4}{0} [4] \oplus [\cc{0_0}]
  + \binom{4}{1} [3] \oplus [\cc{1_0}]
  + \binom{4}{2} [2] \oplus [\cb{0_1},\cc{2_0}]
  + \binom{4}{3} [1] \oplus [\cb{1_1},\cc{3_0}]
  + \binom{4}{4} [0] \oplus [\ca{0_2},\cb{2_1},\cc{4_0}] \tag{i=2} \\
  & \binom{6}{0} [6] \oplus [\cd{0_0}] 
  + \binom{6}{1} [5] \oplus [\cd{1_0}]
  + \binom{6}{2} [4] \oplus [\cc{0_1},\cd{2_0}]
  + \binom{6}{3} [3] \oplus [\cc{1_1},\cd{3_0}]
  + \binom{6}{4} [2] \oplus [\cb{2_2},\cc{3_1},\cd{4_0}]
  + \binom{6}{5} [1] \oplus [\cb{1_2},\cc{3_1},\cd{5_0}]
  + \binom{6}{6} [0] \oplus [\ca{0_3},\cb{2_2},\cc{4_1},\cd{6_0}]  \tag{i=3}
 \end{align}
 The binomial coefficients are a reminder of the permutations induced by the 
 permutation operator $\hat{P}_j^{(2i)}$ acting on $A^{[N](2i)}$.
 The number $[i-j]$ left of $\oplus$ is the degree of the polynomial that is not 
 included in the integration (as it does not contain the integration variable),
 and has the coefficient $A^{[N](2i)}$.
 The numbers $[a,b,\dots]$ are the degrees of the polynomials coming from
 the integral $I_{(j)}^{[N]}$.
 The $\oplus$ means that the orders have to be added, i.e.\ $[1] \oplus [1,3]$
 represents one monomial of order $2$ and one monomial of order $4$ in the 
 final expression.
 The colors of the numbers right of $\oplus$ indicate the same order of the 
 resulting monomial and the subscript indicates the number $m$ of \eqref{eq:iN} 
 from which the monomial is obtained.
 We note that all contributions to $A^{[N-1](0)}$ come from the terms for 
 \cd{$m = i$}, all contributions to $A^{[N-1](2)}$ come from 
 the terms for \cc{$m = i-1$}, etc.
 From this structure, \eqref{eq:aiter} can be obtained as follows:
 First, we change the labels of the summation variables of the coordinates in 
 \eqref{eq:iN} from $k_1,\dots,k_{j-2m}$ to $k_{2i-j+1},\dots,k_{2(i-m)}$,
 \begin{align}
 I_j^{[N]} &= \sum_{m=0}^{\floor*{\frac{j}{2}}}
               \frac{(-1)^j j!}{4^m m! (j-2m)!} 
                \frac{I_0^{[N]}}{\left(S_{NN}^{[N]}\right)^{j-m}} 
                \sum_{k_{i-j+1}=1}^{N-1} \dotsi \sum_{k_{i-2m}=1}^{N-1} 
                 S_{k_{2i-j+1} N}^{[N]} \dots S_{k_{2(i-m)} N}^{[N]}
                 x_{k_{2i-j+1}}\dots x_{k_{2(i-m)}}
 \end{align}
 so that we can insert this formula directly into the equation for the density
 after first integration \eqref{eq:rhonm1},
 \begin{align}
  \rho^{[N-1]}(x) &= 
    \sum_{i=0}^M 
     \sum_{j=0}^{2i} 
      \sum_{m=0}^{\floor*{\frac{j}{2}}}
       \frac{(-1)^j j!}{4^m m! (j-2m)!} 
       \sqrt{\frac{\pi}{S_{NN}^{[N]}}}
       \frac{\Gamma^{[N-1]}}{\left(S_{NN}^{[N]}\right)^{j-m}}
       \sum_{k_1=1}^{N-1} \dotsi \sum_{k_{i-2m}=1}^{N-1} 
        \hat{P}_j^{(2i)} A_{k_1 \dots k_{2i-j} N \dots}^{[N](2i)} 
        S_{k_{2i-j+1} N}^{[N]} \dots S_{k_{2(i-m)} N}^{[N]}  
        x_{k_1} \dots x_{k_{2(i-m)}} \label{eq:l1} \\
   &\stackrel{!}{=} 
        \Gamma^{[N-1]}
        \left( 
        A^{[N-1](0)} + 
        \sum_{k_1,k_2=1}^{N-1} A_{k_1 k_2}^{[N-1](2)} x_{k_1} x_{k_2} + 
        \dots +
        \sum_{k_1,\dots,k_{2M}=1}^{N-1} A_{k_1 \dots k_{2M}}^{[N-1](2M)} x_{k_1} \dots x_{k_{2M}}
        \right) \label{eq:l2}
 \end{align}
 Now we have to identify $A^{[N-1](2\alpha)}$ of \eqref{eq:l2} in \eqref{eq:l1}
 by setting $m = i-\alpha$ and by ensuring that the limits of the sums over 
 $i$ and $j$ are adjusted accordingly.
}}

To obtain the one-nucleus density, equations \eqref{eq:siter} and 
\eqref{eq:aiter} need to be iterated until only three indices are left.
Those belong to the displacement coordinates $x_1, x_2, x_3$.
In order to obtain the one-nucleus density for the other nuclei, the procedure 
is repeated after appropriate permutation of the columns of transformation 
matrix $T_{jk}$.

Last, we note that that we could ignore all factors $1/\sqrt{\pi}$ in the 
original and updated coefficients because each integration cancels one of the 
of the initially $N$ factors.
Then we need to multiply the final one-nucleus density with $\pi^{-3/2}$ to 
obey the normalization condition.

\section{Note on the computational implementation}

The approximations that are used for the nuclear wavefunction allow to compute 
the vibrational one-nucleus densities for molecules with a relatively large number of 
nuclei $N_n$.
However, the resulting polynomial is of order $2M$, where $M$ is the sum of the 
vibrational quanta in the system.
The current numerical implementation stores arrays of the polynomial 
coefficients that are of dimension $N_n^{2M}$, hence with the number of quanta 
$M$ the memory limit is quickly reached, so that in practice only computations 
for $M \le 4$ are possible on a modern workstation.
With a different numerical implementation this problem can possibly be avoided, 
but for large $M$ the local harmonic approximation that is made in the normal 
mode analysis is questionable anyway, hence this is not practical restriction.

\clearpage

\begin{center}
  \Large Supporting Information for ``On the Probability Density of the Nuclei in a Vibrationally Excited Molecule'':
Additional Examples
\end{center}

{\bf A note on the labeling of the normal modes:}
\emph{
In contrast to the main article, molecules with more that three nuclei are 
discussed in the following, hence there are more normal modes.
To have a uniform and simple notation for all molecules, the following 
convention is used to label the normal modes:
The normal modes are numbered according to the frequency of their corresponding 
harmonic oscillators in ascending order.
Modes 1 to 6 are those of translation and rotation of the molecule.
The symmetry of the mode is ignored in the numbering.
}

{\bf A note on the labeling of the states:}
\emph{
In these document, only one-nucleus densities for a single excitation of one 
mode or, for methane, for two singly excited modes or one double excited mode
are shown.
The state is labeled by the number(s) of the excited modes.
}

In the following, present one-nucleus densities for selected states of water,
mono-deterated benzene, ethene, mono-deuterated ethene (D-ethene), and methane
are presented and it is discussed how their qualitative shape can 
be predicted by using the LOCO rules and the normal mode coordinates.
All computations were performed as described in the main article, hence 
the Born-Oppenheimer approximation is used in a local harmonic approximation 
at the nuclear configuration of minimum energy (the equilibrium configuration), 
and it is assumed that the molecule is localized and oriented.
The three-dimensional reference space has coordinates $R_1, R_2, R_3$.
For each molecule, first the normal mode coordinates are discussed and thereafter 
selected densities are presented.

\begin{figure}[htbp]
  \centering
  \includegraphics[width=.99\textwidth]{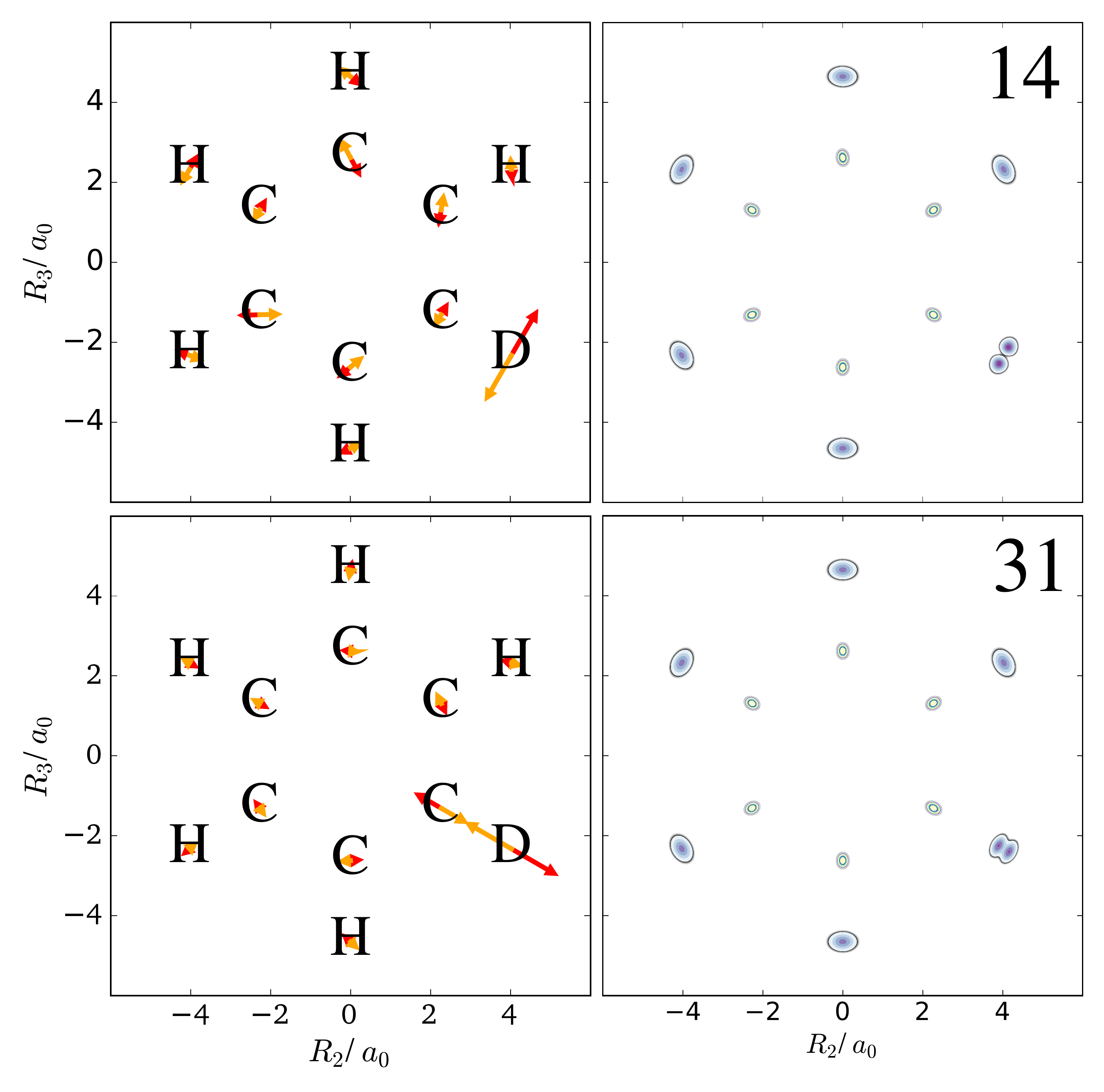}
  \caption{Left: Normal mode coordinates of the mono-deuterated benzene molecule.
  Right: Contour plots of the one-nucleus densities of a localized and oriented 
  mono-deuterated benzene molecule in the molecular plane for vibrational states 
  corresponding to the first excitations of the normal modes shown to the left.
  }
  \label{fig:pic_benzene_deuterated}
\end{figure}

As mentioned in the main article, for the benzene molecule the first excitations of any of 
the normal mode coordinates do not lead to a qualitative change of the 
one-nucleus density compared to the ground state because the hydrogen and oxygen 
nuclei are displaced in the same direction by multiple normal modes.
In contrast, for mono-deterated benzene there are only two normal modes that 
significantly displaced the deuterium nucleus.
These coordinates are also perpendicular with respect to each other, hence their 
excitations are clearly visible in the one-nucleus density.
This is shown in figure \ref{fig:pic_benzene_deuterated}.

\begin{figure}[htbp]
  \centering
  \includegraphics[width=.8\textwidth]{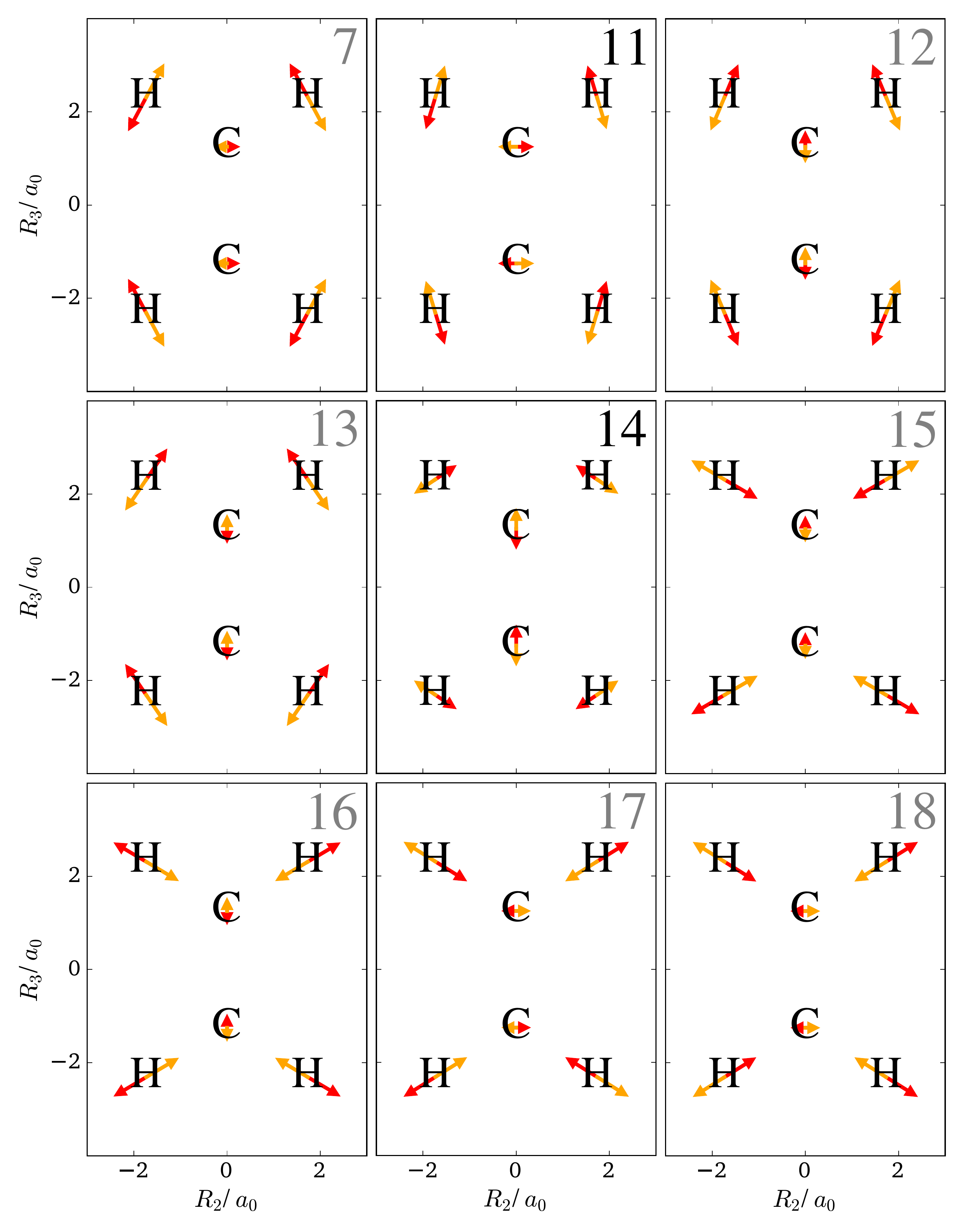}
  \caption{Normal modes of ethene that are confined to the molecular plane.
  The modes are labeled according to frequency, with modes 1-6 representing
  translation and rotation of the whole molecule.
  The arrows show the extent (length) of the displacement of the nuclei along 
  the mode and the directionality of the displacement (color).
  }
  \label{fig:ethene_nm}
\end{figure}

The equilibrium configuration of ethene is planar, hence the normal mode
coordinates correspond to displacements of the nuclei that are either 
completely in the molecular plane, or perpendicular.
Hence, the discussion is restricted to the normal modes in the molecular plane, 
which is defined as the $R_2$-$R_2$ plane.
For ethene, figure \ref{fig:ethene_nm} shows the selected normal mode coordinates.
The modes are numbered by increasing frequency of the corresponding harmonic 
oscillator, and modes 1-6 are those of translation and rotation of the whole system.

From the figure, it can be seen that at the hydrogen nuclei there are many normal 
modes that correspond to displacements in a similar spatial direction, which
also have similar magnitude.
For example, modes 7, 11, 12, and 13 all displace the hydrogen nuclei to a 
similar extend in a similar direction, while modes 15, 16, 17, and 18 do the 
same in an almost perpendicular direction.
Hence, from the LOCO rules it can be concluded that an excitation of just one 
of these modes does not yield any notable qualitative changes of the one-nucleus 
density, i.e.\ there are no new clear minima appearing that correspond to the 
nodes of the wavefunction in such an excited state.
This is indeed the case, and all one-nucleus densities for an excitation along 
one of these modes is qualitatively similar to the ground state.

\begin{figure}[htbp]
  \centering
  \includegraphics[width=.8\textwidth]{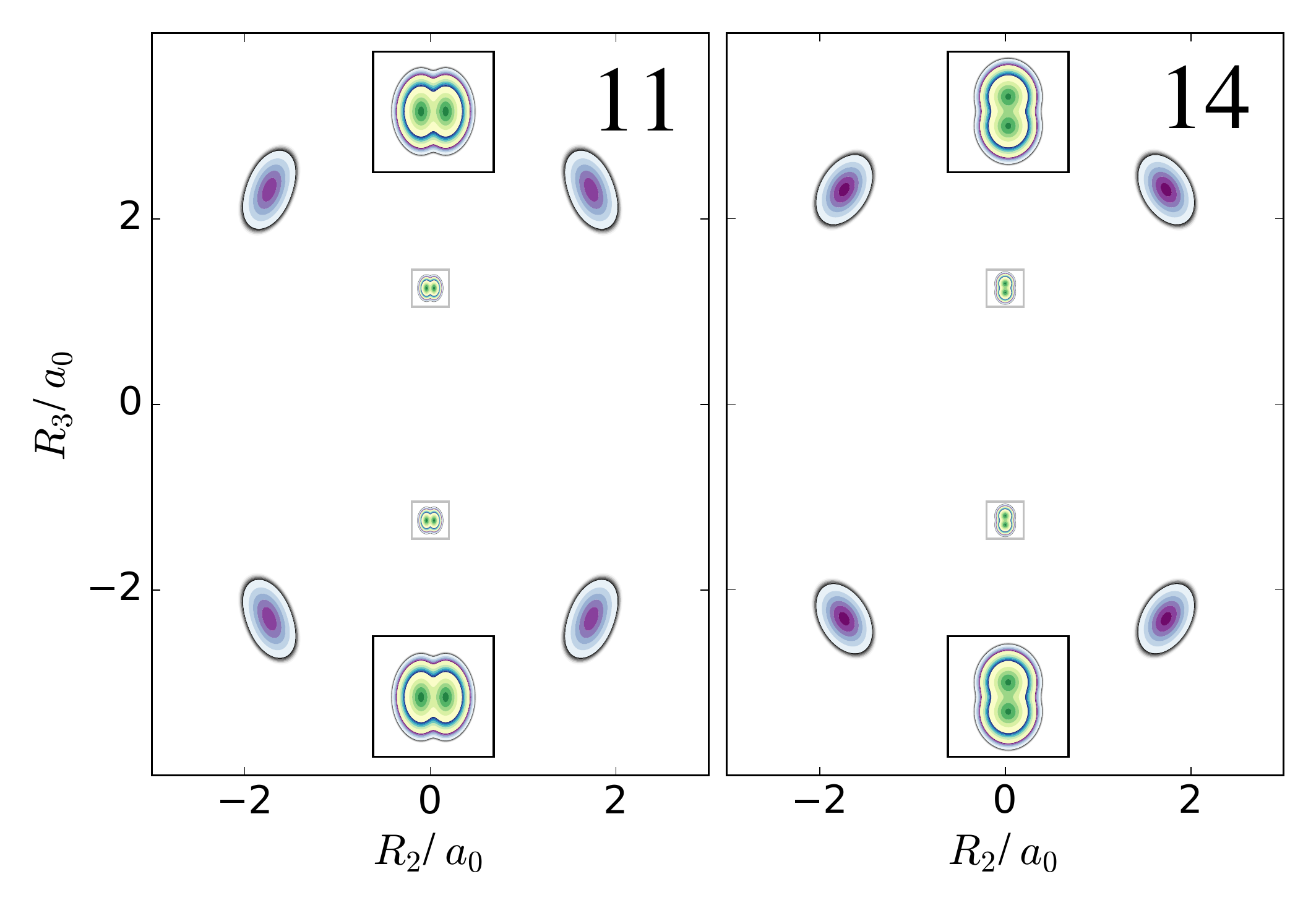}
  \caption{Contour plots of the one-nucleus densities of localized and oriented 
  ethene in the molecular plane for the vibrational states corresponding to the 
  first excitation along normal modes 11 and 14. Insets at the top and bottom 
  show a magnified view of the region around the oxygen nuclei.}
  \label{fig:ethene_dens}
\end{figure}

The situation at the carbon nuclei, however, is somewhat different.
Although there are again many modes that displace these nuclei in the same 
direction, there are two modes that displace them stronger than all other modes:
Mode 11 displaces the oxygen nuclei comparably strongly along $R_2$, while mode
14 displaces the oxygen nuclei comparably strongly along $R_3$.
This difference can be seen in the one-nucleus densities for the respective 
excited states.
Figure \ref{fig:ethene_dens} shows contour plots of the one-nucleus densities for the 
first excited states of mode 11 and mode 14, and insets show a magnification of 
the regions around the carbon nuclei.
Several contour maps with different line spacing are used to be able to see 
details of the density of the hydrogen nuclei and of the carbon nuclei in the 
same picture.
The density at those nuclei has two local maxima and a depletion in the region
of the equilibrium position.
No other of the excited states corresponding to the first excitation any of 
the normal modes shows this qualitative features, neither at the carbon nuclei
nor at the hydrogen nuclei.

\begin{figure}[htbp]
  \centering
  \includegraphics[width=.8\textwidth]{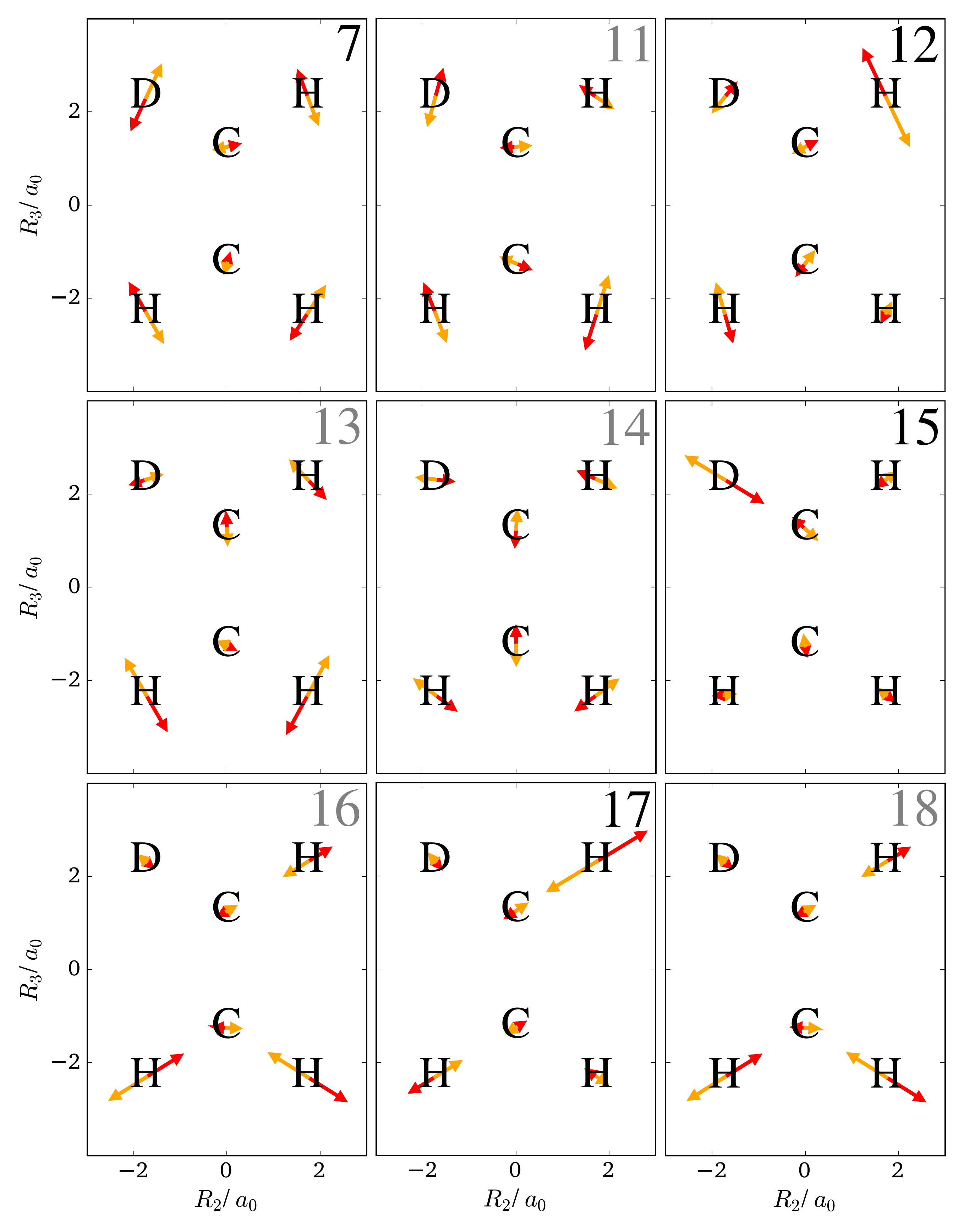}
  \caption{Normal modes of mono-deuterated ethene that are confined to the 
  molecular plane. The modes are labeled according to frequency, with modes 1-6 
  representing translation and rotation of the whole molecule.
  }
  \label{fig:dethene_nm}
\end{figure}

The situation is altered if the symmetry is broken by isotope substitution.
The normal mode coordinates for D-ethene are given in figure \ref{fig:dethene_nm}.
As is clear from the figure, there are several normal modes that should, if 
excited, according to the LOCO rules yield a clear qualitative imprint in the 
one-nucleus density, because they are the only ones that displace the considered
nucleus significantly in a certain direction.
To illustrate, normal modes 7, 12, 15, and 17 are used.
Normal modes 7 and 15 are the ones that displace the deuterium nucleus the 
strongest, and in almost perpendicular directions.
Modes 12 and 17 displace the top-right hydrogen nucleus the strongest, and also
in mutually almost perpendicular directions.
Consequently, excitations of those modes can be expected to have the strongest 
qualitative impact on the one-nucleus density at the given nucleus.

\begin{figure}[htbp]
  \centering
  \includegraphics[width=.8\textwidth]{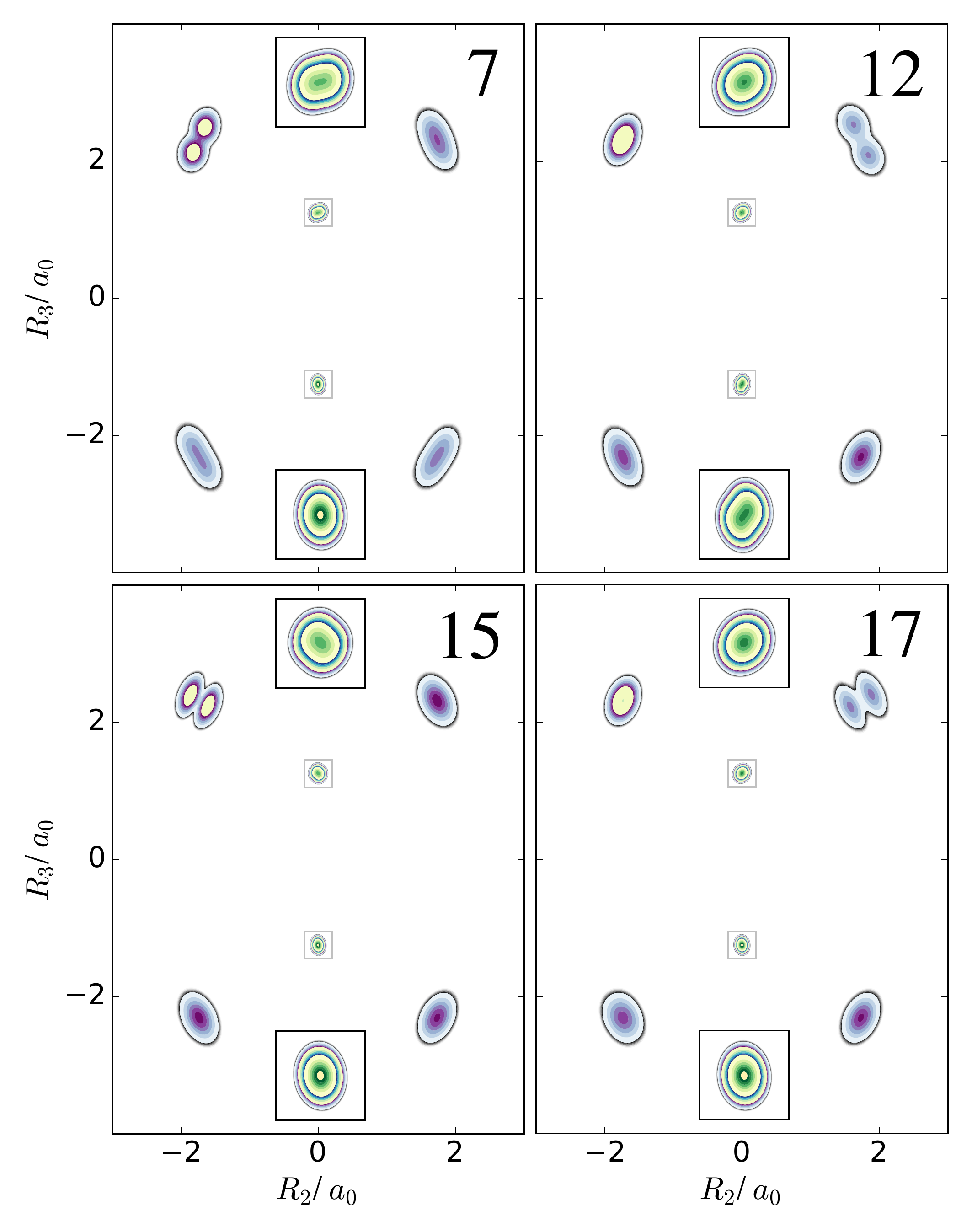}
  \caption{Contour plots of the one-nucleus densities of localized and oriented 
  mono-deuterated ethene in the molecular plane for the vibrational states 
  corresponding to the first excitation along normal modes 7, 12, 15, and 17. 
  Insets at the top and bottom show a magnified view of the region around the 
  oxygen nuclei.}
  \label{fig:dethene_dens}
\end{figure}

Figure \ref{fig:dethene_dens} shows the one-nucleus density for the excites states
represented by first excitation of modes 7, 12, 15, or 17.
It can clearly be seen that excitations of mode 7 and 15 as well as 12 and 17 show 
two maxima and a depletion corresponding to the node in the wavefunction of 
the excited state, at the deuterium nucleus and the top-right hydrogen nucleus,
respectively.
There are more examples of this behavior at the other nuclei, but none of these
is as pronounced as the ones depicted in figure \ref{fig:dethene_dens}.

\begin{figure}[htbp]
  \centering
  \includegraphics[width=.8\textwidth]{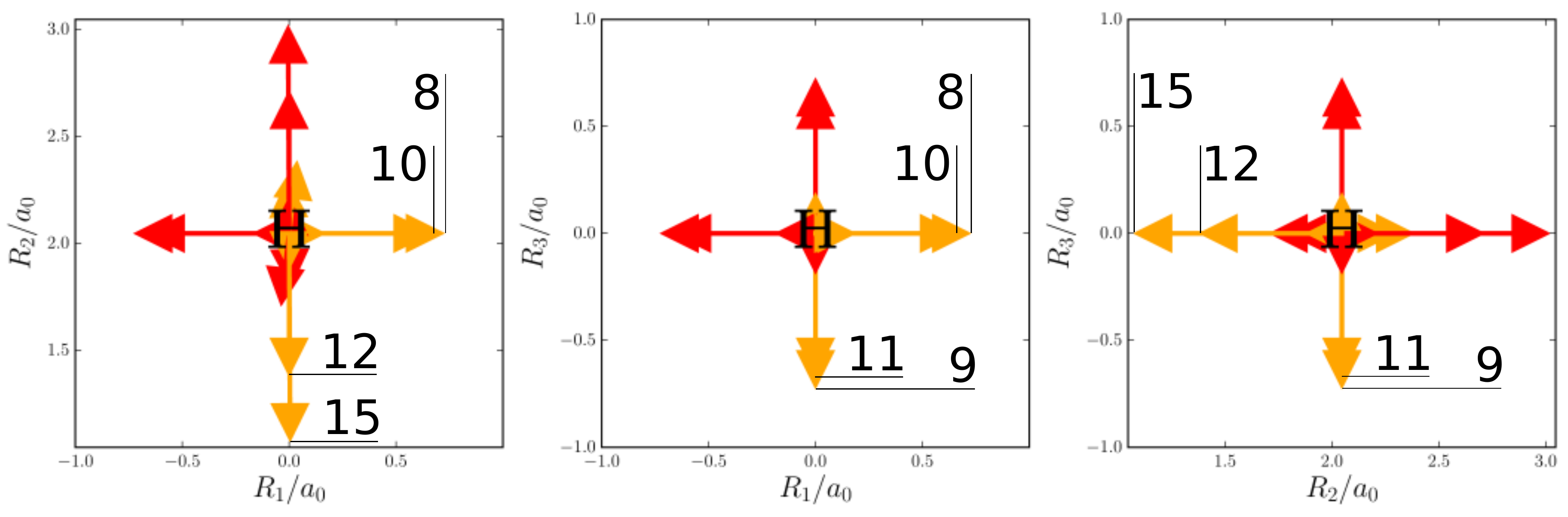}
  \caption{Normal modes of one of the hydrogen nuclei of methane 
  in the $R_1$-$R_2$-, $R_1$-$R_3$-, and $R_2$-$R_3$-plane (the center of mass
  of the molecule is at the origin).
  For those corresponding to the largest displacements in a given direction, 
  their numbers (ordered according to increasing frequency of the normal modes,
  with modes 1-6 corresponding to translation and rotation of the 
  molecule) are given. 
  }
  \label{fig:methane_nm}
\end{figure}

Last, it should be illustrated that the LOCO rules also hold for non-planar 
molecules.
For this purpose, methane is considered.
In its equilibrium configuration, this molecule has four hydrogen nuclei at the 
vertices of a tetrahedron, and a carbon nucleus at its center.
As it is hard to draw the one-nucleus density of the nuclei in a picture, 
only one of the four equivalent hydrogen nuclei is considered.
For the carbon nucleus being at the origin, this nucleus is located at 
ca.\ $(0,2,0) a_0$.
All normal mode coordinates at this nucleus are shown in figure 
\ref{fig:methane_nm} as arrows in the $R_1$-$R_2$-, $R_1$-$R_3$-, and 
$R_2$-$R_3$-plane.

From the figure, it can be seen that at this nucleus, in each direction there 
are only two relevant normal modes:
mode 8 and 10 along $R_1$, mode 12 and 15 along $R_2$, and mode 9 and 11 along
$R_3$.
Mode 8, 15, and 9 are those displacing the nucleus the strongest, although 
only mode 15 has a clear margin with respect to mode 12, while the displacement
along modes 10 and 11 are very close to those of modes 8 and 9, respectively

\begin{figure}[htbp]
  \centering
  \includegraphics[width=.8\textwidth]{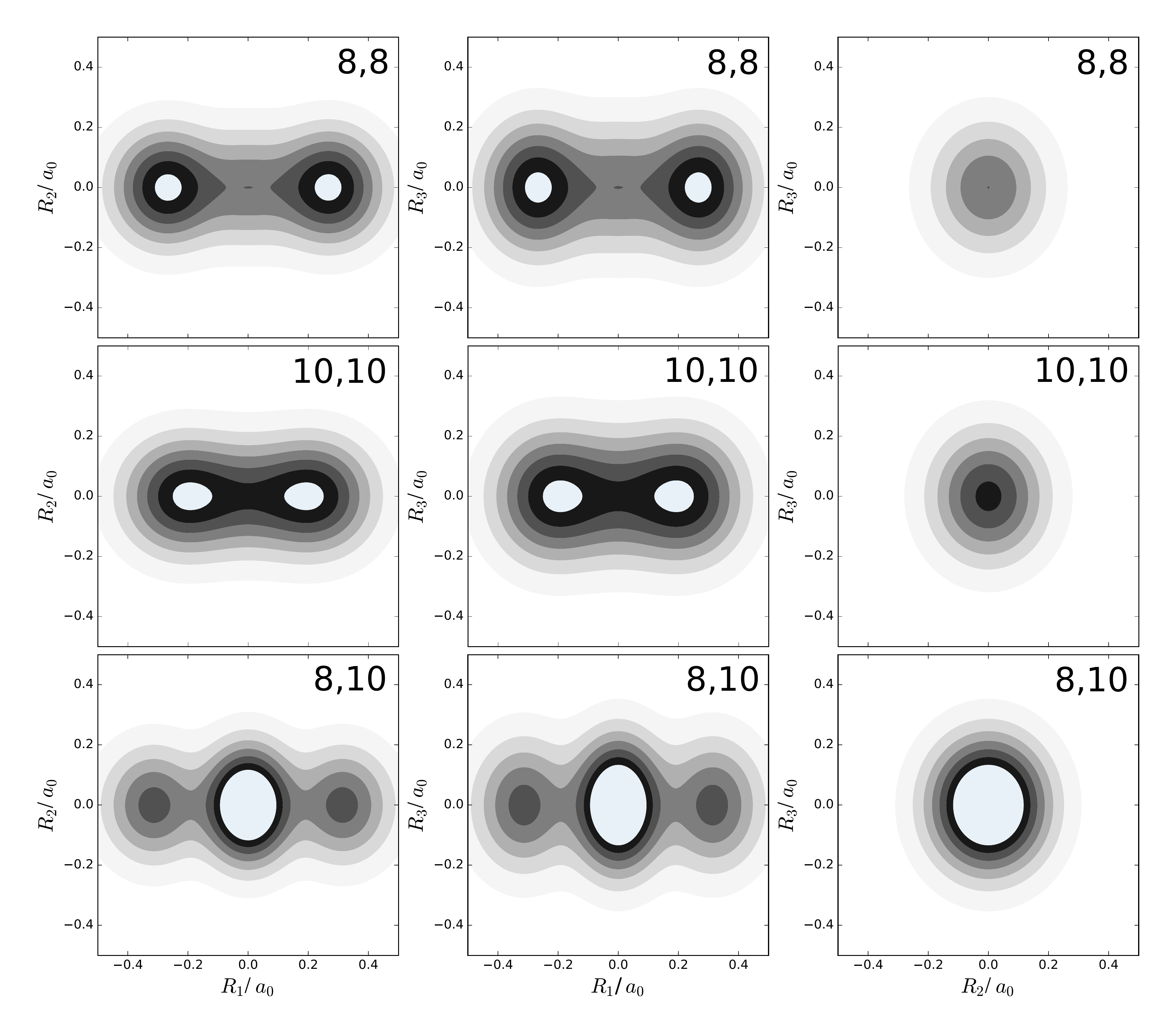}
  \caption{Contour plots of the one-nucleus densities of the hydrogen 
  nucleus of methane of figure \ref{fig:methane_nm} for a state where two 
  quanta are distributed in the normal modes. 
  Left column: $R_1$-$R_2$-plane.
  Middle column: $R_1$-$R_3$-plane. 
  Right column: $R_2$-$R_3$-plane.
  Top row: One-nucleus density for the state corresponding to the second 
  excitation of mode 8.
  Middle row: One-nucleus density for the state corresponding to the second 
  excitation of mode 10.
  Bottom row: One-nucleus density for the state corresponding to the first 
  excitation of both mode 8 and mode 10.
  Note that compared to the normal modes in figure \ref{fig:methane_nm},
  the hydrogen nucleus was shifted to the origin in $R_2$-direction.
  }
  \label{fig:methane_xdens}
\end{figure}

\begin{figure}[htbp]
  \centering
  \includegraphics[width=.8\textwidth]{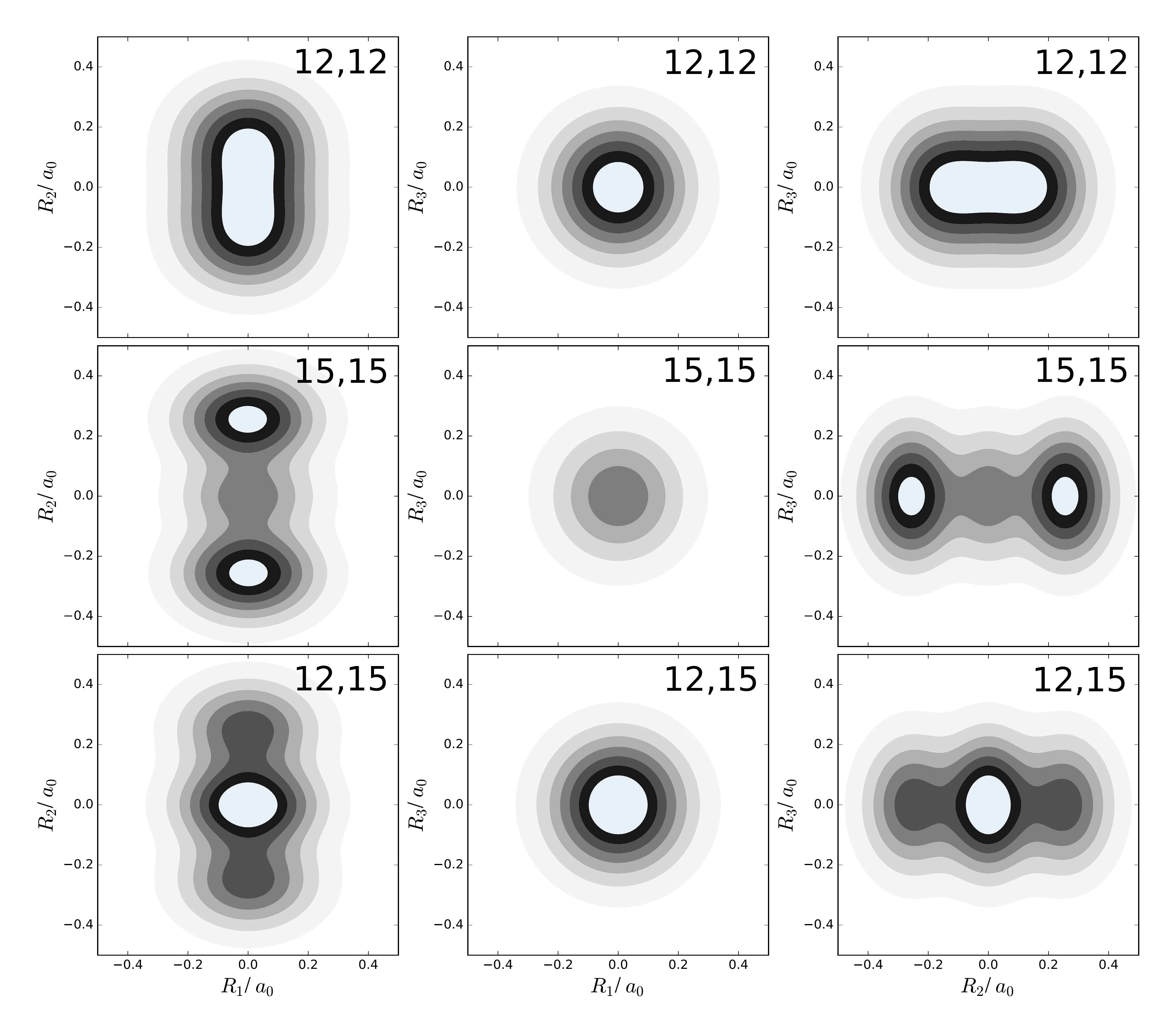}
  \caption{Contour plots of the one-nucleus densities of the hydrogen 
  nucleus of methane of figure \ref{fig:methane_nm} for a state where two 
  quanta are distributed in the normal modes. 
  Left column: $R_1$-$R_2$-plane.
  Middle column: $R_1$-$R_3$-plane. 
  Right column: $R_2$-$R_3$-plane.
  Top row: One-nucleus density for the state corresponding to the second 
  excitation of mode 12.
  Middle row: One-nucleus density for the state corresponding to the second 
  excitation of mode 15.
  Bottom row: One-nucleus density for the state corresponding to the first 
  excitation of both mode 12 and mode 15.
  Note that compared to the normal modes in figure \ref{fig:methane_nm},
  the hydrogen nucleus was shifted to the origin in $R_2$-direction.
  }
  \label{fig:methane_ydens}
\end{figure}

\begin{figure}[htbp]
  \centering
  \includegraphics[width=.8\textwidth]{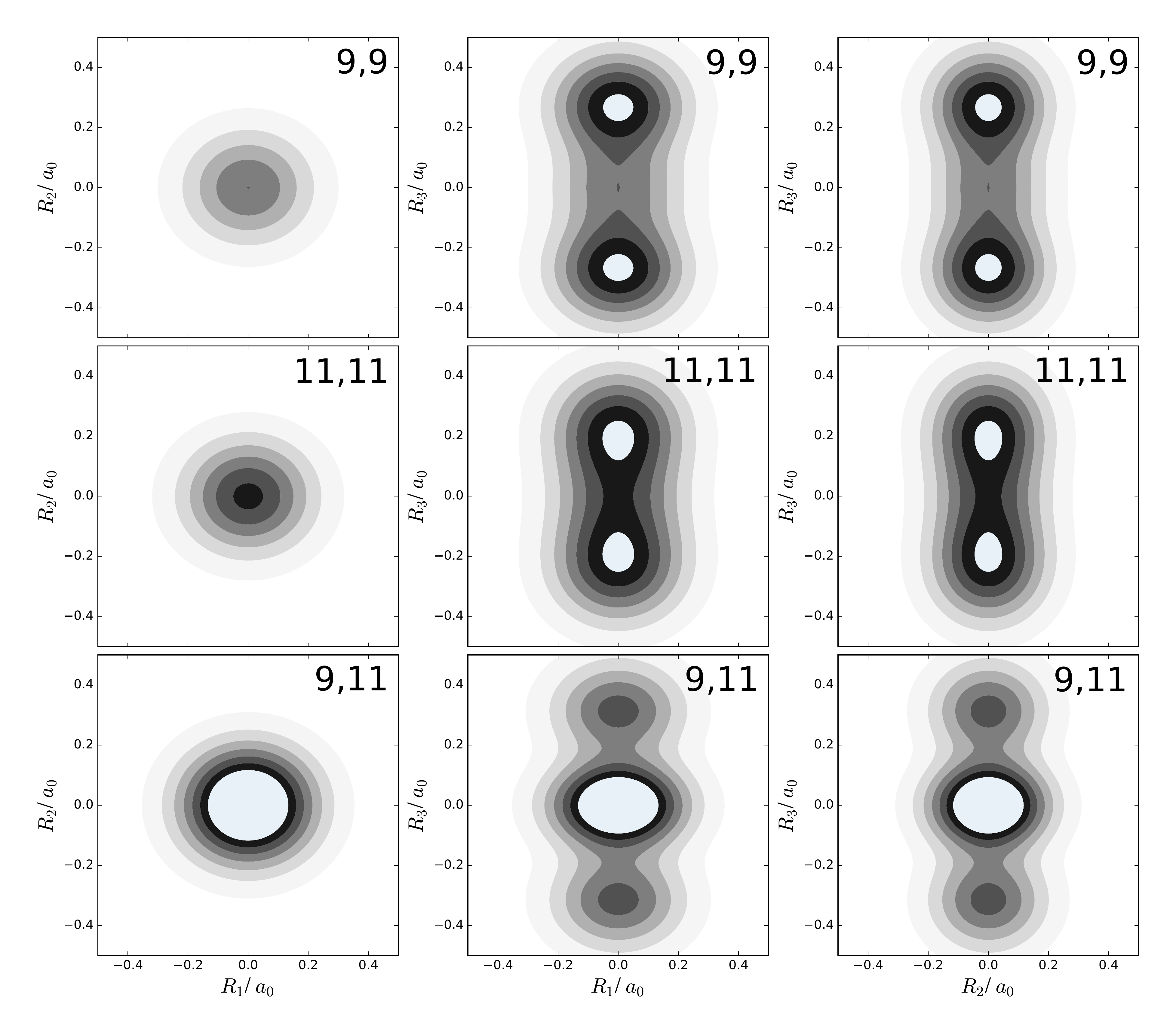}
  \caption{Contour plots of the one-nucleus densities of the hydrogen 
  nucleus of methane of figure \ref{fig:methane_nm} for a state where two 
  quanta are distributed in the normal modes. 
  Left column: $R_1$-$R_2$-plane.
  Middle column: $R_1$-$R_3$-plane. 
  Right column: $R_2$-$R_3$-plane.
  Top row: One-nucleus density for the state corresponding to the second 
  excitation of mode 9.
  Middle row: One-nucleus density for the state corresponding to the second 
  excitation of mode 11.
  Bottom row: One-nucleus density for the state corresponding to the first 
  excitation of both mode 9 and mode 11.
  Note that compared to the normal modes in figure \ref{fig:methane_nm},
  the hydrogen nucleus was shifted to the origin in $R_2$-direction.
  }
  \label{fig:methane_zdens}
\end{figure}

In this example, the excited states corresponding to two quanta in these modes
are considered.
Contour plots of the one-nucleus densities are given in figure \ref{fig:methane_xdens}
for excitations of mode 8 and 10, in figure \ref{fig:methane_ydens}
for excitations of mode 12 and 15, and in figure \ref{fig:methane_zdens}
for excitations of mode 9 and 11.

It is found that a double excitation of mode 15 has a clear triple-maximum 
structure reminiscent of the density of the harmonic oscillator wavefunction
in this node.
According to the LOCO rules this is expected, because mode 15 displaces the 
considered nucleus the most.
Also double excitations of modes 8 or 9 yield triple-maximum structures, 
although the central maximum is almost invisible.
Again, this can be rationalized because the wavefunction is in its ground state
along modes 10 and 11, which have similar effects on the nucleus.

The combined effect of exciting two locally similar modes can also be observed.
In figures \ref{fig:methane_xdens}, \ref{fig:methane_ydens}, and 
\ref{fig:methane_zdens}, the one-nucleus density of the excited state corresponding 
to an excitation of mode 8 and 10, mode 12 and 15, and mode 9 and 11 are also 
shown.
A triple-maximum structure is found, which is most pronounced when the two modes
are locally similar, i.e.\ for excitation of mode 8 and 10 as well as mode 9 
and 11.

\clearpage


\clearpage

\bibliographystyle{unsrt}
\bibliography{lit}

\begin{thebibliography}{10}

\bibitem{sutcliffe05pccp}
Brian~T. Sutcliffe and R.~Guy Woolley.
\newblock {Molecular structure calculations without clamping the nuclei}.
\newblock {\em Phys. Chem. Chem. Phys.}, 7(21):3664, 2005.

\bibitem{sutcliffe10tca}
Brian Sutcliffe.
\newblock {To what question is the clamped-nuclei electronic potential the
  answer?}
\newblock {\em Theor. Chem. Acc.}, 127(3):121, 2010.

\bibitem{matyus11pra}
Edit M\'atyus, J\"urg Hutter, Ulrich M\"uller-Herold, and Markus Reiher.
\newblock {On the emergence of molecular structure}.
\newblock {\em Phys. Rev. A}, 83(5):052512, 2011.

\bibitem{matyus11jcp}
Edit M\'atyus, J\"urg Hutter, Ulrich M\"uller-Herold, and Markus Reiher.
\newblock {Extracting elements of molecular structure from the all-particle
  wave function}.
\newblock {\em J. Chem. Phys}, 135(20), 2011.

\bibitem{matyus12jcp}
Edit M\'atyus and Markus Reiher.
\newblock {Molecular structure calculations: A unified quantum mechanical
  description of electrons and nuclei using explicitly correlated Gaussian
  functions and the global vector representation}.
\newblock {\em J. Chem. Phys}, 137(2), 2012.

\bibitem{manz2014}
J\"orn Manz, Jhon~Fredy P\'erez-Torres, and Yonggang Yang.
\newblock {Vibrating H$_2^+$($^{2}\Sigma_{\rm g}^+$, $JM = 00$) Ion as a
  Pulsating Quantum Bubble in the Laboratory Frame}.
\newblock {\em J. Phys. Chem. A}, 118:8411, 2014.

\bibitem{bredtmann2015}
Timm Bredtmann, Dennis~J. Diestler, Si-Dian Li, J\"orn Manz, Jhon~Fredy
  P\'erez-Torres, Wen-Juan Tian, Yan-Bo Wu, Yonggang Yang, and Hua-Jin Zhai.
\newblock {Quantum theory of concerted electronic and nuclear fluxes associated
  with adiabatic intramolecular processes}.
\newblock {\em Phys. Chem. Chem. Phys.}, 17:29421--29464, 2015.

\bibitem{torres2015}
Jhon~Fredy P\'erez-Torres.
\newblock {Dissociating H$_2^+$($^{2}\Sigma_{\rm g}^+$,$JM=00$) Ion as an
  Exploding Quantum Bubble}.
\newblock {\em J. Phys. Chem. A}, 119:2895, 2015.

\bibitem{diestler2018}
Dennis~J. Diestler, D.~Jia, J.~Manz, and Y.~Yang.
\newblock {Na$_2$ Vibrating in the Double-Well Potential of State
  $2^{1}\Sigma_{\rm u}^+$ ($JM = 00$): A Pulsating "Quantum Bubble" with
  Antagonistic Electronic Flux}.
\newblock {\em J. Phys. Chem. A}, 122:2150, 2018.

\bibitem{abedi12}
Ali Abedi, Neepa~T. Maitra, and E.~K.~U. Gross.
\newblock {Correlated electron-nuclear dynamics: Exact factorization of the
  molecular wavefunction}.
\newblock {\em J. Chem. Phys.}, 137:22A530, 2012.

\bibitem{born1927adp}
M.~Born and R.~Oppenheimer.
\newblock {Zur Quantentheorie der Molekeln}.
\newblock {\em Annalen der Physik}, 389(20):457, 1927.

\bibitem{schild2016}
Axel Schild, Federica Agostini, and E.~K.~U. Gross.
\newblock {Electronic Flux Density beyond the Born-Oppenheimer Approximation}.
\newblock {\em J. Phys. Chem. A}, 120(19):3316--3325, 2016.

\bibitem{smit2001}
Michael~J. Smit, Gerrit~C. Groenenboom, Paul E.~S. Wormer, Ad~van~der Avoird,
  Robert Bukowski, and Krzysztof Szalewicz.
\newblock {Vibrations, Tunneling, and Transition Dipole Moments in the Water
  Dimer}.
\newblock {\em J. Phys. Chem. A}, 105(25):6212, 2001.

\bibitem{dawes13}
Richard Dawes, Xiao-Gang Wang, and Tucker Carrington.
\newblock {CO Dimer: New Potential Energy Surface and Rovibrational
  Calculations}.
\newblock {\em J. Phys. Chem. A}, 117:7612, 2013.

\bibitem{welsch15}
Ralph Welsch and Uwe Manthe.
\newblock {Full-dimensional and reduced-dimensional calculations of initial
  state-selected reaction probabilities studying the H + CH$_4$ $\rightarrow$
  H$_2$ + CH$_3$ reaction on a neural network PES}.
\newblock {\em J. Chem. Phys}, 142(6):064309, 2015.

\bibitem{donoghue2016}
Geoff Donoghue, Xiao-Gang Wang, Richard Dawes, and Tucker Carrington.
\newblock {Computational study of the rovibrational spectra of
  CO$_2$-C$_2$H$_2$ and CO$_2$-C$_2$D$_2$}.
\newblock {\em Journal of Molecular Spectroscopy}, 330:170, 2016.

\bibitem{shapiro1981}
Moshe Shapiro.
\newblock {Photofragmentation and mapping of nuclear wavefunctions}.
\newblock {\em Chemical Physics Letters}, 81(3):521, 1981.

\bibitem{zewail00}
Ahmed~H. Zewail.
\newblock {Femtochemistry: Atomic-Scale Dynamics of the Chemical Bond}.
\newblock {\em J. Phys. Chem. A}, 104(24):5660, 2000.

\bibitem{jurek2004}
Z.~Jurek, G.~Oszl{\'{a}}nyi, and G.~Faigel.
\newblock {Imaging atom clusters by hard X-ray free-electron lasers}.
\newblock {\em Europhysics Letters ({EPL})}, 65(4):491, 2004.

\bibitem{ergler2006}
Th. Ergler, A.~Rudenko, B.~Feuerstein, K.~Zrost, C.~D. Schr\"oter,
  R.~Moshammer, and J.~Ullrich.
\newblock {Spatiotemporal Imaging of Ultrafast Molecular Motion: Collapse and
  Revival of the $\mathrm{D}_{2}{}^{+}$ Nuclear Wave Packet}.
\newblock {\em Phys. Rev. Lett.}, 97:193001, 2006.

\bibitem{schmidt2012}
L.~Ph.~H. Schmidt, T.~Jahnke, A.~Czasch, M.~Sch\"offler, H.~Schmidt-B\"ocking,
  and R.~D\"orner.
\newblock {Spatial Imaging of the H$_{2}{}^{+}$ Vibrational Wave Function at
  the Quantum Limit}.
\newblock {\em Phys. Rev. Lett.}, 108:073202, 2012.

\bibitem{kimberg2014}
Victor Kimberg and Catalin Miron.
\newblock {Molecular potentials and wave function mapping by high-resolution
  electron spectroscopy and ab initio calculations}.
\newblock {\em Journal of Electron Spectroscopy and Related Phenomena},
  195:301, 2014.

\bibitem{zeller2016}
Stefan Zeller, Maksim Kunitski, J{\"o}rg Voigtsberger, Anton Kalinin, Alexander
  Schottelius, Carl Schober, Markus Waitz, Hendrik Sann, Alexander Hartung,
  Tobias Bauer, Martin Pitzer, Florian Trinter, Christoph Goihl, Christian
  Janke, Martin Richter, Gregor Kastirke, Miriam Weller, Achim Czasch, Markus
  Kitzler, Markus Braune, Robert~E. Grisenti, Wieland Sch{\"o}llkopf, Lothar
  Ph.~H. Schmidt, Markus~S. Sch{\"o}ffler, Joshua~B. Williams, Till Jahnke, and
  Reinhard D{\"o}rner.
\newblock {Imaging the He$_2$ quantum halo state using a free electron laser}.
\newblock {\em Proceedings of the National Academy of Sciences}, 113(51):14651,
  2016.

\bibitem{manz13prl}
J\"orn Manz, Jhon~Fredy P\'erez-Torres, and Yonggang Yang.
\newblock {Nuclear Fluxes in Diatomic Molecules Deduced from Pump-Probe Spectra
  with Spatiotemporal Resolutions down to 5 pm and 200 asec}.
\newblock {\em Phys. Rev. Lett.}, 111:153004, 2013.

\bibitem{barth15}
I.~Barth, C.~Daniel, E.~Gindensperger, J.~Manz, J.~F. P\'erez-Torres,
  A.~Schild, C.~Stemmle, D.~Sulzer, and Y.~Yang.
\newblock {Intramolecular Nuclear Flux Densities}.
\newblock In S.~H. Lin, A.~A. Villaeys, and Y.~Fujimura, editors, {\em Advances
  in Multi-Photon Processes and Spectroscopy}, volume~22, page~59. World
  Scientific, Singapore, 2015.

\bibitem{bredtmann2015pccp}
Timm Bredtmann, Dennis~J. Diestler, Si-Dian Li, J\"orn Manz, Jhon~Fredy
  P\'erez-Torres, Wen-Juan Tian, Yan-Bo Wu, Yonggang Yang, and Hua-Jin Zhai.
\newblock {Quantum theory of concerted electronic and nuclear fluxes associated
  with adiabatic intramolecular processes}.
\newblock {\em Phys. Chem. Chem. Phys.}, 17(44):29421--29464, 2015.

\bibitem{waitz2017}
M.~Waitz, R.~Y. Bello, D.~Metz, J.~Lower, F.~Trinter, C.~Schober, M.~Keiling,
  U.~Lenz, M.~Pitzer, K.~Mertens, M.~Martins, J.~Viefhaus, T.~Weber S.~Klumpp,
  L.~Ph.~H. Schmidt, J.~B. Williams, M.~S. Schöffler, V.~V. Serov, A.~S.
  Kheifets, L.~Argenti, A.~Palacios, F.~Mart\'in, T.~Jahnke, and R.~D\"orner.
\newblock {Imaging the square of the correlated two-electron wave function of a
  hydrogen molecule}.
\newblock {\em Nature Communications}, 8:2266, 2017.

\bibitem{ullrich2012}
Carsten~A. Ullrich.
\newblock {\em {Time-Dependent Density-Functional Theory}}.
\newblock Oxford University Press, Oxford, United Kingdom, 2012.

\bibitem{moore2008}
Amanda~M. Moore and Paul~S. Weiss.
\newblock {Functional and Spectroscopic Measurements with Scanning Tunneling
  Microscopy}.
\newblock {\em Annual Review of Analytical Chemistry}, 1(1):857, 2008.

\bibitem{baer2010}
Roi Baer.
\newblock {Ground-State Degeneracies Leave Recognizable Topological Scars in
  the Electronic Density}.
\newblock {\em Phys. Rev. Lett.}, 104:073001, 2010.

\bibitem{mccarron2018}
Daniel McCarron.
\newblock {Laser cooling and trapping molecules}.
\newblock {\em J. Phys. B: At. Mol. Opt. Phys.}, 51:212001, 2018.

\bibitem{henriksen2011}
Niels~Engholm Henriksen and Flemming~Yssing Hansen.
\newblock {\em {Theories of Molecular Reaction Dynamics: The Microscopic
  Foundation of Chemical Kinetics (Oxford Graduate Texts)}}.
\newblock Oxford University Press, 2011.

\bibitem{eckart1935pra}
Carl Eckart.
\newblock {Some Studies Concerning Rotating Axes and Polyatomic Molecules}.
\newblock {\em Phys. Rev.}, 47(7):552, 1935.

\bibitem{littlejohn1997rmp}
Robert~G. Littlejohn and Matthias Reinsch.
\newblock {Gauge fields in the separation of rotations and internal motions in
  the n-body problem}.
\newblock {\em Rev. Mod. Phys.}, 69(1):213, 1997.

\bibitem{bunker2006}
Philip~R. Bunker and Per Jensen.
\newblock {\em {Molecular symmetry and spectroscopy}}.
\newblock NRC Research Press, Ottawa, 2006.

\bibitem{lauvergnat2016jcp}
David Lauvergnat, Josep~M. Luis, Bernard Kirtman, Heribert Reis, and Andr\'e
  Nauts.
\newblock {Numerical and exact kinetic energy operator using Eckart conditions
  with one or several reference geometries: Application to HONO}.
\newblock {\em J. Chem. Phys}, 144(8), 2016.

\bibitem{turney12}
Justin~M. Turney, Andrew~C. Simmonett, Robert~M. Parrish, Edward~G. Hohenstein,
  Francesco~A. Evangelista, Justin~T. Fermann, Benjamin~J. Mintz, Lori~A.
  Burns, Jeremiah~J. Wilke, Micah~L. Abrams, Nicholas~J. Russ, Matthew~L.
  Leininger, Curtis~L. Janssen, Edward~T. Seidl, Wesley~D. Allen, Henry~F.
  Schaefer, Rollin~A. King, Edward~F. Valeev, C.~David Sherrill, and T.~Daniel
  Crawford.
\newblock {Psi4: an open-source ab initio electronic structure program}.
\newblock {\em Wiley Interdiscip. Rev. Comput. Mol. Sci.}, 2(4):556, 2012.

\bibitem{dunning1989jcp}
Thom~H. Dunning.
\newblock {Gaussian basis sets for use in correlated molecular calculations. I.
  The atoms boron through neon and hydrogen}.
\newblock {\em J. Chem. Phys}, 90(2):1007--1023, 1989.

\bibitem{shimanouchi1972}
T.~Shimanouchi.
\newblock {\em {Tables of Molecular Vibrational Frequencies Consolidated Volume
  I}}.
\newblock National Bureau of Standards, 1972.

\bibitem{braun1993}
Charles~L. Braun and Sergei~N. Smirnov.
\newblock {Why is water blue?}
\newblock {\em Journal of Chemical Education}, 70(8):612, 1993.

\bibitem{cohentannoudji2007}
Claude Cohen-Tannoudji, Bernard Diu, and Frank Lalo\"e.
\newblock {\em Quantenmechanik 1/2}.
\newblock de Gruyter, Berlin, 2007.

\bibitem{jeffrey2008}
Alan Jeffrey and Hui-Hui Dai.
\newblock {\em {Handbook of Mathematical Formulas and Integrals}}.
\newblock Elsevier, 2008.

\end{thebibliography}

\end{document}